\documentclass[onecolumn,authoryear]{els-mrw}

\usepackage{amsmath,amssymb,amsfonts,amsthm,makeidx,graphicx}
\usepackage{txfonts}
\usepackage{helvet}
\usepackage{microtype}

\newcommand \Pomeron {I\!\!P}

\begin{document}

\chapter{Nuclear parton distributions and nuclear shadowing}

\author[1]{Petja Paakkinen}
\author[2,3]{Vadim Guzey}

\address[1]{\orgname{CERN}, \orgdiv{Theoretical Physics Department}, \orgaddress{1211 Geneva 23, Switzerland}}
\address[2]{\orgname{University of Jyväskylä}, \orgdiv{Department of Physics}, \orgaddress{P.O. Box 35, 40014 University of Jyväskylä, Finland}}
\address[3]{\orgname{Helsinki Institute of Physics}, \orgaddress{P.O. Box 64, 00014 University of Helsinki, Finland}}

\maketitle

\begin{abstract}[Abstract]

In this contribution, we review the physics and phenomenology of nuclear parton distribution functions (PDFs), with a particular focus on the small-$x$ region of nuclear shadowing. We summarise experimental evidence for nuclear modifications of PDFs and discuss their theoretical explanations; in particular, we contrast different models of nuclear shadowing. We also overview model-agnostic extractions of nuclear PDFs from global data on hard processes with nuclei, emphasizing the validity of collinear factorization and the dominance of leading-twist nuclear PDFs. Finally, we discuss perspectives for present and future precision studies of nuclear PDFs and small-$x$ QCD dynamics, including photonuclear reactions in heavy-ion ultraperipheral collisions at the Large Hadron Collider and lepton-nucleus deep inelastic scattering at the future Electron-Ion Collider.

\end{abstract}

\section{Introduction}
\label{sec:intro}

Modern high-energy nuclear physics explores Quantum Chromodynamics (QCD) in strongly interacting matter and, in particular, studies the subatomic structure of atomic nuclei in terms of their most fundamental constituents: quarks and gluons (collectively called partons). The partonic picture of nuclei is encoded in nuclear parton distribution functions (PDFs), and their precise mapping constitutes an essential part of the physics programs at the Relativistic Heavy Ion Collider (RHIC), the Large Hadron Collider (LHC), and the future Electron-Ion Collider (EIC). In addition to providing first-principles theoretical framework for the description and interpretation of hard processes involving nuclei, nuclear PDFs play an important role in defining the initial conditions for heavy-ion collisions at the RHIC and LHC that lead to the creation of novel states of matter such as the quark–gluon plasma (QGP), in particular for the production of the hard probes that traverse the QGP and can reveal its properties. The nuclear PDFs find also applications in predictions of ultra-high-energy neutrino interactions with matter and the production of prompt atmospheric leptons and neutrinos, both relevant for measurements at neutrino observatories.

Nuclear PDFs are markedly different from those of free nucleons and, while the pattern of their nuclear modifications is generally well established through QCD analyses of data on many observables from multiple experiments, their origin admits different, sometimes complementary or conflicting, theoretical explanations. One of the most actively explored is the regime of small momentum fraction $x$, where nuclear PDFs are suppressed compared to their free-nucleon counterparts because of the phenomenon known as nuclear shadowing. Encoding these nuclear effects into the nuclear PDFs generally assumes the applicability of collinear factorization and dominance of single-parton scattering (leading twist) contributions with linear scale evolution of the parton distributions. This interpretation is backed up by the experimental verification that nuclear-PDF based predictions work very well for a wide array of data and observables in the region where the perturbative expansion of QCD (pQCD) is applicable. In this view, the analyses of nuclear PDFs also work as a baseline for the searches of new, nonlinear small-$x$ QCD effects such as parton saturation.

In this contribution, we overview nuclear PDFs in the framework of collinear factorization of QCD, with an emphasis on the small-$x$ region of nuclear shadowing. After presenting formal definitions of nuclear PDFs, their properties, and connection to nucleon PDFs in Sec.~\ref{sec:factorization}, we summarize experimental evidence for nuclear modifications in PDFs in Sec.~\ref{sec:experiment}. Theoretical models offering interpretation of nuclear PDFs at large $x$ are outlined in Sec.~\ref{sec:large-x}, while approaches to small-$x$ shadowing, including the Gribov-Glaber theory, the leading twist approximation, the dipole model, Color Glass Condensate (CGC)-based approaches, and calculations based on resummation of nucleus-enhanced higher-twist contributions, are reviewed in Sec.~\ref{sec:small-x}. Another central part is a critical digest of global analysis approaches to nuclear PDFs in Sec.~\ref{sec:global-analysis}. Section~\ref{sec:summary} presents the summary and outlook.

\section{Basic definitions and collinear factorization in nuclei}
\label{sec:factorization}

Nuclear PDFs, like any other parton distribution functions, are defined in the context of QCD collinear factorization~\citep{Collins:1989gx}. This factorization enables separating the physics of short-distance, large momentum transfer, partonic scattering processes that are calculable from first principles in pQCD and the long-distance, partonic structure of hadrons and nuclei, which is non-perturbative in nature. In a general form, this allows one to write the cross section for the production of a hard inclusive final state $k+X$ in a collision of two particles $A$ and $B$ with momenta $P_A$, $P_B$ as
\begin{equation}
  \sigma^{A+B \rightarrow k+X} (P_A,P_B,Q^2) \!\overset{Q \gg \Lambda_{\rm QCD}}{=}\! \sum_{i,j,X'} \int_0^1 \mathrm{d}x_A \int_0^1 \mathrm{d}x_B \, f_i^A(x_A,Q^2) \, f_j^B(x_B,Q^2) \, \hat{\sigma}^{i+j \rightarrow k+X'}(x_AP_A,x_BP_B,Q^2) + \mathcal{O}(1/Q^2),
  \label{eq:coll-fact}
\end{equation}
where by \emph{hard} we mean that the process should involve a large momentum-trasfer scale $Q \gg \Lambda_{\rm QCD}$ and \emph{inclusive} refers to the requirement that we allow for any number and configuration of additional particles $X$ to be produced together with the hard probe $k$ in this process. In this expression, the functions $f_i^A(x_A,Q^2)$ are the parton distribution functions, which describe the distribution of partons of flavour $i$ carrying a fraction $x_A$ of the momentum $P_A$ in the particle $A$, when probed at the resolution scale $Q^2$; the PDFs $f_j^B(x_B,Q^2)$ have a similar interpretation with respect to particle $B$. These PDFs are convolved with the partonic cross sections, $\hat{\sigma}^{i+j \rightarrow k+X'}(x_AP_A,x_BP_B,Q^2)$, and summed over all possible initial-state flavours $i,j$ and partonic final states $k + X'$ to obtain the hard-process cross section. (It is understood that $X$ is the sum of $X'$ plus the beam/target remnants.) Where necessary, i.e.\ when one is describing the inclusive production of a hadron $h$ in the final state, the partonic cross section needs to be further convolved with parton-to-hadron fragmentation functions $D_k^h(z,Q^2)$, with $z$ the fraction of outgoing parton momentum carried by the hadron. At high enough momentum-transfer scale, any additional contributions $\mathcal{O}(1/Q^2)$ should be negligible. These power corrections contain target-mass corrections (see~\cite{Ruiz:2023ozv}) as well as contributions from multi-parton interactions. Borrowing from the language of operator-product expansion, the latter are often referred to as higher-twist effects, and correspondingly the functions $f_i^A$ as leading-twist PDFs.

The framework of nuclear PDFs applies the collinear factorization approach to collisions where either $A$ or $B$ (or both of them) is a nucleus. The nuclear state thus plays no special role at this level; the nucleus is simply treated as a collection of partons similarly as one would do with factorising collisions involving hadrons (protons, neutrons, pions, etc.). Correspondingly, the scale evolution of nuclear PDFs obeys the same DGLAP equations~\citep{Dokshitzer:1977sg,Gribov:1972ri,Lipatov:1974qm,Altarelli:1977zs} as their hadronic counterparts. It is only when one tries to interpret or predict the nuclear PDFs in terms of nucleonic (proton and neutron) degrees of freedom that understanding the associated nuclear physics becomes necessary. (But this is exactly what one would like to achieve!) Like for hadron PDFs, it is also possible to give a field-theoretic operator definition for the nuclear parton distributions. Following the definition in~\cite{Collins:1981uw}, this reads
\begin{equation}
  f_i^A(x_A, Q^2) = \int \frac{\mathrm{d}z^-}{4\pi} \mathrm{e}^{\mathrm{i}x_AP_A^+z^-} \left.\! \langle A,Z,P_A | \, \bar{\psi}_q(z)\gamma^+\mathcal{W}(z,0)\psi_q(0) \, | A,Z,P_A \rangle \, \right|_{z^+=z_\perp=0} \quad \text{for}\ i=q.
  \label{eq:pdf-def}
\end{equation}
Here, $| A,Z,P_A \rangle$ is the nuclear state with a mass number $A$ and a charge $Z$ that is moving along the $z$-direction with a four-momentum $P_A$ (we do not keep track of the spin state). The QCD operator $\bar{\psi}_q(z)\gamma^+\mathcal{W}(z,0)\psi_q(0)$, which is understood to be renormalised at the hard scale $Q$, consists of a quark-field operators $\psi_q$ and its conjugate, a Dirac gamma matrix $\gamma^+$, and a Wilson line $\mathcal{W}$ over a light-like distance $z$. For understanding how these factors come into play and for the definition of the gluon PDF, we refer the reader to the textbook by~\cite{Collins:2011zzd} and the corresponding chapter in this Encyclopedia.

\subsection{Connecting the parton distributions of nuclear and nucleonic degrees of freedom}

The function $f_i^A(x_A, Q^2)$ above depends on the fraction $x_A \in [0,1]$ of the total momentum $P_A$ carried by the entire nucleus. It turns out to be useful to describe the nuclear PDFs in terms of a scaled variable $x_N = A x_A \in [0,A]$ (where the subscript $N$ is often dropped for brevity), which now refers to a fraction of the \emph{average nucleon momentum} $P_A / A$ carried by the parton. In fact, nuclear PDFs are typically given in terms of $x_N$ distributions
\begin{equation}
  f_i^A(x_N, Q^2) = \frac{1}{A}f_i^A(x_A = x_N/A, Q^2),
\end{equation}
which are more convenient for direct comparison between different nuclei and with free-nucleon PDFs. (The scaling by $1/A$ is needed in order to preserve a sensible normalisations of the integral sum rules in Sec.~\ref{sec:sum_rules}.) In a simple picture where nucleons carry approximately equal shares of the nuclear momentum, values $x_N > 1$ are kinematically suppressed, and one can in most applications neglect the contribution from this region (small effects can arise from Fermi motion and short-range correlations, see Sec.~\ref{sec:large-x}). In particular, the DGLAP equations for nuclear PDFs in terms of the $x_N$ variable read
\begin{equation}
  \frac{\mathrm{d}f_i^A(x_N, Q^2)}{\mathrm{d}\ln Q^2} = \frac{\alpha_s(Q^2)}{2\pi} \int_{x_N}^A \, \frac{\mathrm{d}z_N}{z_N} P_{ij}(x_N/z_N, \alpha_s(Q^2)) \, f^A_j(z_N,Q^2),
  \label{eq:dglap}
\end{equation}
where $\alpha_s$ is the strong coupling and $P_{ij}$ the partonic splitting functions, but by assuming that $f_i^A(x_N > 1, Q^2) = 0$, the upper integration limit can be set to $z_N < 1$, and Eq.~\eqref{eq:dglap} becomes identical to the free-nucleon evolution equations.

Because nuclei contain both protons and neutrons, nuclear PDFs depend on both the mass number $A$ and the atomic number $Z$. It is therefore common to decompose them as
\begin{equation}
  f^A_i(x_N, Q^2) = Z \, f^{p/A}_i(x_N, Q^2) + (A-Z) \, f^{n/A}_i(x_N, Q^2),
  \label{eq:full-nucleus-from-bound-nucleons}
\end{equation}
where $f^{p/A}_i$ and $f^{n/A}_i$ are the bound-proton and bound-neutron PDFs in a nucleus $A$. This kind of a decomposition helps distinguishing isospin effects, which are caused by different numbers of protons and neutrons being present in nuclei, from ``genuine'' nuclear effects that cause bound nucleon PDFs to differ from the free nucleon ones. The latter are often described in terms of nuclear modification ratios of the bound nucleons ($N=p,n$),
\begin{equation}
  R^{N/A}_i(x_N, Q^2) = \frac{f^{N/A}_i(x_N, Q^2)}{f^N_i(x_N, Q^2)},
  \label{eq:nucl-mod-bound-nucleon}
\end{equation}
where $f^N_i(x_N, Q^2)$ is the PDF of a free nucleon. Similarly, one can define
\begin{equation}
  R^{A}_i(x_N, Q^2) = \frac{f^{A}_i(x_N, Q^2)}{Z \, f^{p}_i(x_N, Q^2) + (A-Z) \, f^{n}_i(x_N, Q^2)}
  \label{eq:nucl-mod-full-nucleus}
\end{equation}
for the nuclear modification of the entire nucleus. At the limit of no nuclear effects beyond isospin, these ratios would collapse to exactly one at all $x_N$, $Q^2$. Assuming isospin symmetry (see Sec.~\ref{sec:isospin_symmetry}), the two ratios are equal to each other for flavours other than up and down (anti-)quarks.

As will become clear later, the bound-nucleon PDFs and the corresponding nuclear modification ratios need to be understood as effective quantities, averaging over contributions from various nuclear configurations. In fact, by decomposing the nuclear PDFs as in Eq.~\eqref{eq:full-nucleus-from-bound-nucleons}, one is possibly including also non-nucleonic degrees of freedom of the nucleus wave function in the effective bound-nucleon PDFs (see Sec.~\ref{sec:large-x}), and in the small-$x_N$ shadowing region, it is not even possible to distinguish a scattering off an individual nucleon from a coherent scattering off a collection of them (see Sec.~\ref{sec:small-x}).

\subsection{Sum rules}
\label{sec:sum_rules}

Since QCD conserves charge and flavour quantum numbers, the parton distributions obey associated sum rules. For nuclear PDFs, these are often expressed in terms of the bound-nucleon PDFs. Notably, the baryon-number and charge sum rules for the entire nucleus,
\begin{equation}
  \frac{1}{3} \sum_q \int_0^1 \mathrm{d}x_A \, \left(f^A_q(x_A, Q^2) - f^A_{\bar{q}}(x_A, Q^2)\right) = A, \qquad \sum_q \int_0^1 \mathrm{d}x_A \, e_q \left(f^A_q(x_A, Q^2) - f^A_{\bar{q}}(x_A, Q^2)\right) = Z,
\end{equation}
where $e_q$ is the charge of a quark with flavour $q$, as well as the requirements for net-zero total strangeness, charm and bottom flavour number, are fulfilled by requiring bound-nucleon quark number sum rules
\begin{equation}
\begin{split}
  \int_0^A \mathrm{d}x_N \, \left(f^{N/A}_u(x_N, Q^2) - f^{N/A}_{\bar{u}}(x_N, Q^2)\right) = \begin{cases}
    2 & \text{for}\ N=p \\
    1 & \text{for}\ N=n
  \end{cases}, \qquad&
  \int_0^A \mathrm{d}x_N \, \left(f^{N/A}_d(x_N, Q^2) - f^{N/A}_{\bar{d}}(x_N, Q^2)\right) = \begin{cases}
    1 & \text{for}\ N=p \\
    2 & \text{for}\ N=n
  \end{cases}, \\
  \int_0^A \mathrm{d}x_N \, \left(f^{N/A}_q(x_N, Q^2) - f^{N/A}_{\bar{q}}(x_N, Q^2)\right) &= 0 \quad \text{for}\ N \in \{p,n\}\ \text{and}\ q \neq u,d,
\end{split}
\end{equation}
just as one does for free-nucleon PDFs. Similarly, the momentum conservation leads to the momentum sum rule
\begin{equation}
  \sum_i \int_0^1 \mathrm{d}x_A \, x_A \, f^A_i(x_A, Q^2) = \frac{1}{A} \sum_i \int_0^A \mathrm{d}x_N \, x_N \, f^A_i(x_N, Q^2) = \sum_i \int_0^A \mathrm{d}x_N \, x_N \, f^{N/A}_i(x_N, Q^2) = 1,
\end{equation}
where the sum runs over all flavours, including both (anti-)quarks and gluons.

\subsection{Isospin symmetry}
\label{sec:isospin_symmetry}

Due to the close to degenerate masses of the up and down quarks, and the flavour blindness of the strong interaction, QCD has an approximate global SU(2) symmetry called isospin symmetry. This property can be used to impose relations between PDFs of different targets. This can be understood by considering the definition in Eq.~\eqref{eq:pdf-def}, and noting that it is possible to construct a unitary operator $U$ that exchanges up and down quarks at each point in space. Acting on any field or state, this operator transforms a member of an isospin multiplet to a member with opposite isospin charge, times a possible phase factor. In particular, since this transformation is global, in a nuclear state it transforms each proton to neutron and neutron to proton. By inserting identity operators $I = U^\dagger U$ into the parton-distribution definition in Eq.~\eqref{eq:pdf-def}, one then sees that
\begin{equation}
  \langle A,Z,P_A | \, \bar{\psi}_u(z)\gamma^+\mathcal{W}(z,0)\psi_u(0) \, | A,Z,P_A \rangle = \langle A,Z'\!=\!A-Z,P_A | \,  \bar{\psi}_d(z)\gamma^+\mathcal{W}(z,0)\psi_d(0) \, | A,Z'\!=\!A-Z,P_A \rangle,
\end{equation}
i.e.\ the up-quark distribution of a nucleus is equal to the down-quark distribution of its mirror nucleus. The same relation holds also for the up and down antiquarks, whereas for other quark flavours and gluons, which are isospin singlets, a similar consideration tells that the parton distributions of these flavours should be identical for mirror nuclei. As an immediate consequence of the above, for any isoscalar nucleus, i.e.\ $Z = A/2$, the up and down quark distributions are identical up to isospin breaking effects.

Adopting the bound-nucleon decomposition in Eq.~\eqref{eq:full-nucleus-from-bound-nucleons}, and by arguing that the genuine nuclear effects should depend dominantly only on the mass number and not on the nuclear charge, the isospin symmetry is often imposed by assuming a more strict requirement that
\begin{equation}
  \begin{aligned}
    f^{p/A}_u(x_N,Q^2) = f^{n/A}_d(x_N,Q^2), \\
    f^{p/A}_{\bar{u}}(x_N,Q^2) = f^{n/A}_{\bar{d}}(x_N,Q^2),
  \end{aligned} \qquad
  \begin{aligned}
  f^{p/A}_d(x_N,Q^2) = f^{n/A}_u(x_N,Q^2), \\
  f^{p/A}_{\bar{d}}(x_N,Q^2) = f^{n/A}_{\bar{u}}(x_N,Q^2),
  \end{aligned} \qquad
  f^{p/A}_i(x_N,Q^2) = f^{n/A}_i(x_N,Q^2) \quad \text{for}\ i \neq u,\bar{u},d,\bar{d}
  \label{eq:bound-nucleon-isospin}
\end{equation}
in any nucleus $A$. It is good to remember that the isospin symmetry is only approximate, and assumptions made using it must eventually break down. This is easy to understand by considering the nuclear chart: for many stable nuclei the corresponding mirror nucleus does not exist as a bound state or will decay radioactively. In particular, the hydrogen nucleus is a free proton, which has not been observed to decay, whereas its mirror nucleus would be a free neutron, which has a mean lifetime of about 15 minutes. Despite these limitations, the isospin symmetry is very accurate and useful for analyses of nuclear PDFs, and using the relations in Eq.~\eqref{eq:bound-nucleon-isospin} reduces the number of independent functions involved and thus simplifies the problem significantly.

\section{Experimental evidence for nuclear modifications in PDFs}
\label{sec:experiment}

There is now indisputable evidence that the effective parton distributions of bound nucleons are non-trivially modified compared to the free-nucleon ones. These nuclear effects appear consistently across several different processes, signaling that they are due to universal modifications of the PDFs. In the following, we give a brief overview of the available experimental data from different hard-process measurements.

\subsection{Deep inelastic scattering and Drell-Yan dilepton production off nuclear targets}

Much of the understanding about nuclear effects in PDFs originates from deep inelastic scattering (DIS) and Drell-Yan (DY) dilepton production measurements at fixed-target experiments. As purely electromagnetic processes at leading order in pQCD, these processes offer a clean probe of the quark content of nuclei. Historically, these measurements have played an important role in identifying the nuclear effects discussed in Secs.~\ref{sec:large-x} and~\ref{sec:small-x}, and continue to provide the experimental constraints used for the nuclear PDF global analyses discussed in Sec.~\ref{sec:global-analysis}.

\subsubsection{Inclusive $l^\pm+A$ DIS}

The early evidence for nuclear modifications of parton distributions in bound nucleons came from fixed-target charged-lepton DIS experiments ($l^\pm+A \rightarrow l^\pm+X$). In particular, the discovery of the \emph{EMC effect}, named after its first observation at the European Muon Collaboration experiment~\citep{EuropeanMuon:1983wih}, led to the realisation that the nuclear PDFs cannot be described simply as a combination of quasi-free, unmodified nucleon contributions. This first observation was later improved upon and expanded to wider kinematical regions and variety of different nuclei in further experiments at CERN, SLAC, and FNAL, see reviews by \cite{Arneodo:1992wf,Geesaman:1995yd,Norton:2003cb} and references therein, and more recently at JLab~\citep{Seely:2009gt,CLAS:2019vsb,Arrington:2021vuu,HallC:2022utd}. These data are typically given in terms of structure function ratios $R^A_{F_2}(x,Q^2) = F_2^A(x,Q^2) / F_2^d(x,Q^2)$ as a function of the Bjorken-$x$ variable (the momentum transfer $Q^2$ is positively correlated with $x$ in these fixed-target experiments). At leading order in pQCD, this experimentally measurable $x$ is equivalent to the momentum fraction $x_N$ up to kinematical corrections from quark and target mass. For higher orders, this direct relation becomes a convolution due to radiative QCD corrections.

\begin{figure}[t]
\centering
\includegraphics[height=5.2cm]{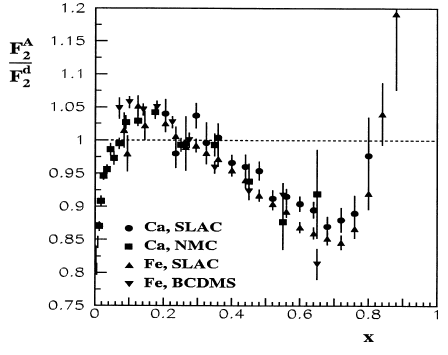}
\hspace{0.25cm}
\includegraphics[height=5.2cm]{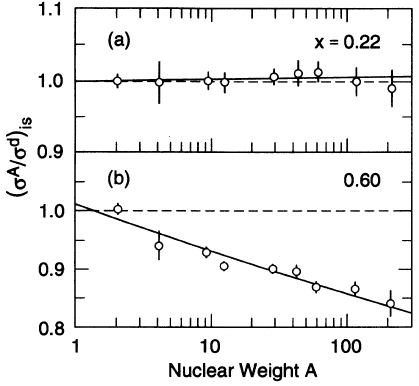}
\caption{(Left) Measurements of structure function ratios for calsium and iron with respect to deuteron as a function of $x$. Reprinted from~\cite{Piller:1999wx}, with permission from Elsevier. (Right) SLAC measurement of the $A$ dependence of DIS cross section ratio to deuteron. Reprinted figure with permission from~\cite{Gomez:1993ri}. Copyright (1994) by the American Physical Society.}
\label{fig:dis}
\end{figure}

Figure~\ref{fig:dis} gives a representative example of these measurements. On the left, we show a compilation by~\cite{Piller:1999wx} of measurements for the calcium and iron to deuteron ratios measured in~\cite{BCDMS:1987upi,Gomez:1993ri,NewMuon:1995cua}. These ratios show the typical shape of nuclear modifications seen in the DIS experiments: At large $x > 0.7$, the ratio rises due to Fermi-motion smearing, for $0.3 < x < 0.7$ we see the suppression by the EMC effect, at $0.1 < x < 0.3$ there is an enhancement known as antishadowing, and finally for $x < 0.1$ one enters the region of nuclear shadowing. On the right, the measurement of the nuclear-mass dependence from~\cite{Gomez:1993ri} is shown. The nuclear effects are observed to grow more or less smoothly as the nuclear mass number $A$ increases. However, for more recent JLab measurements with light nuclei, deviations from the general trend have been observed, and the evolution as a function of $A$ is not necessarily purely monotonic. Note that the shown measurements have been isospin-corrected, as is common for presenting the nuclear DIS data, to correspond for each nucleus to a measurement with a target having the same mass number $A$ but equal numbers of protons and neutrons, with $Z = A/2$. These corrections aid the direct comparisons of cross sections for different nuclei, but have to be properly taken into account or undone in theoretical analyses.

\subsubsection{Inclusive and dimuon $\nu+A$ DIS}
\label{sec:neutrino-dis}

Besides charged-lepton DIS, important complementary information on nuclear parton distributions is obtained from inclusive and dimuon-production neutrino–nucleus DIS ($\nu+A \rightarrow l^\pm(l^\mp)+X$). For these processes, the neutrino–nucleus interaction proceeds via charged-current weak interaction mediated by $W^\pm$ bosons, allowing access to different combinations of quark flavours than in the charged-lepton neutral-current DIS. Inclusive neutrino-induced DIS measurements have been performed at the CDHSW~\citep{Berge:1989hr}, CCFR~\citep{CCFRNuTeV:2000qwc}, NuTeV~\citep{NuTeV:2005wsg}, and CHORUS~\citep{CHORUS:2005cpn} experiments, using heavy nuclear targets of iron and lead. These measurements are available only in terms of absolute cross sections or structure functions, and thus linking them  with nuclear modifications of PDFs has a stronger dependence on what is assumed as the baseline free-nucleon PDFs. Nevertheless, the measurements show in general similar nuclear modifications as the charged-lepton DIS data. Interpreting these measurements has proven somewhat tricky, and some analyses have found tensions across the neutrino data or against charged-lepton measurements, sparking even discussion about possible non-universality (i.e.\ process dependence) of the nuclear modifications, see the discussion in~\cite{Kopeliovich:2012kw}. A significant part of the discrepancies appear to originate from different normalisations of the neutrino-induced data~\citep{Paukkunen:2013grz}, but some tensions are still observed in modern-day analyses, particularly in the small-$x$ shadowing region~\citep{Muzakka:2022wey}.

An important addition to the inclusive data are the measurements of dimuon production in neutrino DIS. Here, the dominant production channel involves a partonic scattering process $\nu+s \rightarrow \mu + c$. The produced charm quark then hadronises, and the produced charmed hadron subsequently decays semileptonically into a muon of opposite charge to that produced in the hard partonic scattering. For discussion on the theoretical description and factorization of this process, see~\cite{Helenius:2024fow}. Having an initial-state strange quark in the leading partonic subprocess makes the dimuon DIS a unique probe of the nucleon strangeness. In fact, the experimental data available from CCFR~\citep{CCFR:1994ikl}, NuTeV~\citep{NuTeV:2001dfo,NuTeV:2007uwm}, and NOMAD~\citep{NOMAD:2013hbk} using again heavy nuclear targets have been traditionally included in analyses of free-proton PDFs in order to constrain the otherwise poorly determined strange-quark PDF, see e.g.~\cite{Faura:2020oom}. The resulting free-nucleon PDFs then ultimately depend on what is assumed for the nuclear modifications in this process, causing a correlation between free-nucleon and nuclear PDF analyses.

\subsubsection{DY in fixed-target $p+A$ and $\pi^\pm+A$ experiments}

Fixed-target DY measurements with proton beams on nuclear targets ($p+A \rightarrow l^\pm l^\mp+X$) were historically important in establishing that the nuclear effects were not specific to DIS, but a more universal feature present in hard interactions with nuclei, as well as for separating the modifications of valence and sea quarks. The measured cross section ratios~\citep{Alde:1990im,NuSea:1999egr} exhibit very clear nuclear shadowing, similar in size to DIS measurements, but no antishadowing, which appears to be nonexistent or significantly less prominent in the DY data. The latter is seen as direct evidence that antishadowing is present only in valence-quark distributions, but not in the sea antiquarks. While important for the shadowing and antishadowing regions, the kinematics of these measurements limit the direct nuclear-PDF sensitivity to $0.01 < x_N < 0.3$ and do not provide significant information on the modifications in the EMC-effect region.

Further DY measurements with nuclear targets have been performed using charged-pion beams ($\pi^\pm+A \rightarrow l^\pm l^\mp+X$). The presence of valence anti-quarks in pions makes this process an interesting probe for the flavour dependence in nuclear PDFs~\citep{Dutta:2010pg,Paakkinen:2016wxk}. Importantly, by taking cross-section ratios, the dependence on pion PDFs can be cancelled to a very good degree. The available data~\citep{NA3:1981yaj,NA10:1987hho,Heinrich:1989cp} probe the nuclear PDFs in a $0.1 < x_N < 0.4$ window and are compatible with the EMC effect observed in DIS measurements, but come with sizeable experimental uncertainties, and do not allow making strong conclusions about the flavour dependence.

\subsection{Hadron colliders – RHIC and LHC}

While the fixed-target DIS and DY experiments gave compelling evidence of the nuclear modifications of quark PDFs, they offered only weak hints on how the gluon distribution in nuclei behaves compared to free nucleons. This was to be changed by the access to novel data types with the high-energy hadron colliders capable of colliding also heavy nuclei, RHIC at BNL and LHC at CERN. The data taking with proton-nucleus and deuteron-nucleus ($p+A$, $d+A$) collisions were motivated in particular as benchmarks for providing a reliable cold-nuclear-matter baseline for studies of QGP properties in nucleus-nucleus ($A+A$) collisions~\citep{Salgado:2011wc,Apolinario:2022vzg}. Importantly, no clear indication of any hot-QCD energy-loss has been found in the $p+A$, $d+A$ collision systems. With the higher center-of-mass energy available in these collisions, especially with the LHC, one was also able to probe partons carrying smaller fractions of momenta and higher momentum transfers than accessible at the fixed-target experiments, expanding the kinematical reach in the nuclear PDF studies by several orders of magnitude in both $x$ and $Q^2$, see~\cite{Klasen:2023uqj} for an illustration of the probed range.

\subsubsection{Hadron and jet production in $p+A$ collisions}

\begin{figure}[t]
\centering
\includegraphics[height=4.85cm]{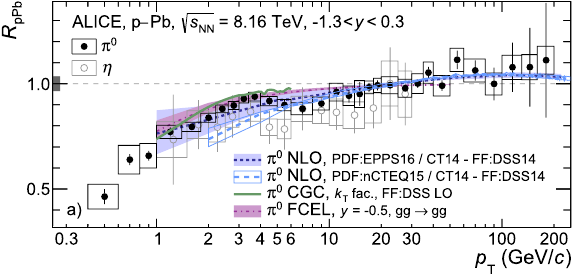}
\includegraphics[height=4.85cm]{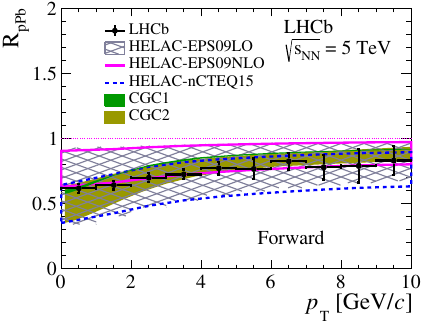}
\caption{(Left) ALICE measurement of $\pi^0$ nuclear modification factor in $p+\mathrm{Pb}$ collisions compared with EPPS16~\citep{Eskola:2016oht} and nCTEQ15~\citep{Kovarik:2015cma} parametrisations of nuclear PDFs. Adapted from~\cite{ALICE:2021est} under CC BY 4.0. (Right) LHCb measurement of D$^0$ nuclear modification factor at a large forward rapidity compared with EPS09~\citep{Eskola:2009uj} and nCTEQ15~\citep{Kovarik:2015cma} parametrisations of nuclear PDFs. Reprinted from~\cite{LHCb:2017yua} under CC BY 4.0. The plots show also comparisons to Colour Glass Condensate (CGC) and fully coherent energy loss (FCEL) based calculations, see Sec.~\ref{sec:global-analysis-vs-higher-twist} for discussion.}
\label{fig:inclhadron}
\end{figure}

Being a QCD process at leading order in perturbation theory, the production of hadrons at large transverse momentum $p_\mathrm{T}$ in collisions of hadrons with nuclei can be used as an effective probe of the gluon distributions of the initial state particles. First sufficinetly high-$p_\mathrm{T}$ data for this process arrived from RHIC $d+\mathrm{Au}$ collisions, and in particular the PHENIX~\citep{PHENIX:2003qdw,PHENIX:2006mhb} and STAR~\citep{STAR:2006xud,STAR:2009qzv} measurements of $\pi^0$ production were historically important in providing first direct experimental evidence for presence of antishadowing in nuclear gluon PDFs as well as for the onset of small-$x_N$ shadowing. These measurements were later extended to other collision systems at RHIC~\citep{PHENIX:2021dod}, and expanded in kinematical reach by the introduction of LHC $p+\mathrm{Pb}$ data, where the ALICE~\citep{ALICE:2016dei,ALICE:2018vhm,ALICE:2021est} and CMS~\citep{CMS:2016xef} mid-rapidity data go deeper in the shadowing region than respective measurements at RHIC due to the larger center-of-mass energy and show a clear suppression in the $R_\mathrm{pPb} = \sigma^{p+\mathrm{Pb}}/\sigma^{p+p}$ nuclear modification factor, see Fig.~\ref{fig:inclhadron} (left). Assuming that hadronisation dynamics are not altered in $p+A$ collisions compared to $p+p$ case, the dependence on fragmentation functions effectively cancels in this ratio. LHCb has also measured the nuclear modification factors for $\pi^0$~\citep{LHCb:2022tjh} and e.g.\ D$^0$~\citep{LHCb:2017yua,LHCb:2022dmh} heavy-flavour production. They find strong suppression in the forward (i.e.\ proton-going, positive rapidity in the LHC convention) direction, see Fig.~\ref{fig:inclhadron} (right), which probes very small $x_N$ of the nucleus, and some moderate enhancement in the backward (nucleus-going, negative rapidity) direction, again in agreement with with expectations from nuclear shadowing and antishadowing, respectively. Similar measurements with production of unidentified (charged) hadrons have been performed, yielding generally the same picture, but hinting towards an enhancement in baryon production relative to light mesons at backward $p+\mathrm{Pb}$ kinematics compared to $p+p$ collisions. In principle, nuclear gluons are probed also in quarkonia measurements in $p+A$ collisions, but their interpretation is convoluted with uncertainties in the production mechanism (see, however,~\cite{Kusina:2017gkz,Duwentaster:2022kpv}). The role of possible higher-twist effects in these processes are discussed in Sec.~\ref{sec:global-analysis-vs-higher-twist}.

\begin{figure}[t]
\centering
\includegraphics[height=5.2cm]{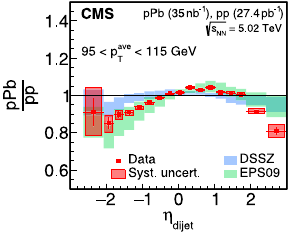}
\hspace{0.25cm}
\includegraphics[height=5.2cm]{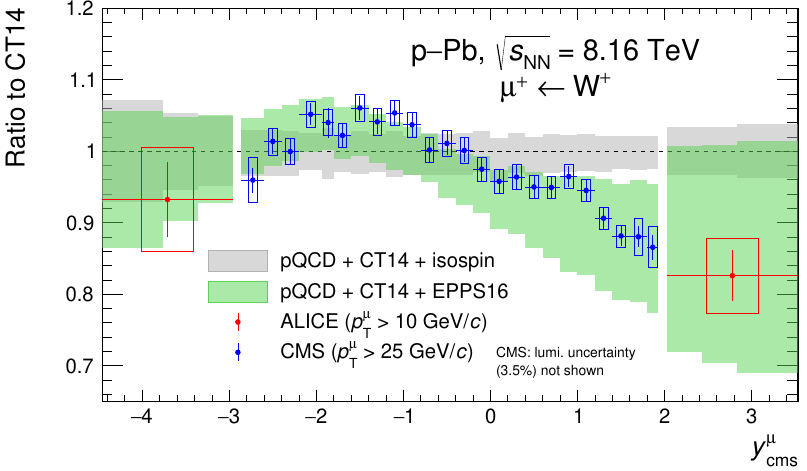}
\caption{(Left) CMS measurement of dijet self-normalised nuclear modification factor in $p+\mathrm{Pb}$ collisions compared with EPS09~\citep{Eskola:2009uj} and DSSZ~\citep{deFlorian:2011fp} parametrisations of nuclear PDFs. Adapted from~\cite{CMS:2018jpl} under CC BY 4.0. (Right) Ratio of ALICE and CMS measurements of $W^+$-boson production with respect to a baseline using only isospin effects, compared with EPPS16~\citep{Eskola:2016oht} parametrisation of nuclear PDFs. Reprinted from~\cite{ALICE:2022cxs} under CC BY 4.0.}
\label{fig:jet_ewboson}
\end{figure}

Instead of considering a single hadron in the final state, one can measure the production of high-$p_\mathrm{T}$ jets, defined through a suitable jet reconstruction algorithm. (Mapping the measured hadron-jet observables to partonic jets calculated in pQCD requires taking non-perturbative hadronisation and underlying-event corrections into account on either side.) In particular, studying the production of a pair of jets, a dijet, gives a good access to the kinematics of the initial-state partons. The CMS measurement of nuclear modification factor for the self-normalised dijet yields as a function of the dijet pseudorapidity $\eta_\mathrm{dijet}$~\citep{CMS:2018jpl}, shown in Fig.~\ref{fig:jet_ewboson} (left), displays a suppression at negative values of $\eta_\mathrm{dijet}$ (EMC-effect), enhancement at midrapidity (antishadowing), and again a substantial suppression at large $\eta_\mathrm{dijet}$ (shadowing). Despite some challenges in the theoretical description of the individual $p+p$ and $p+\mathrm{Pb}$ dijet spectra~\citep{Eskola:2019dui}, these data give again compelling evidence of the gluon PDF nuclear modifications. Currently, there are not many other data sets available for this process, though ATLAS has measured a per-trigger normalised ratio of dijet conditional yields~\citep{ATLAS:2019jgo}, with results that also appear to support the picture of shadowing in nuclear PDFs~\citep{Perepelitsa:2025qpz}.

\subsubsection{Electroweak-boson and $t\bar{t}$ production in $p+A$ and $A+A$ collisions}

The production of heavy electroweak gauge bosons, $W^\pm$ and $Z$, is nowadays considered as important ``standard candle'' processes for hadron colliders as their leptonic decay products are largely insensitive to hadronisation and underlying-event effects. Consequently, the measurements of these processes in $p+\mathrm{Pb}$ collisions, in particular with the improved statistics from LHC Run 2~\citep{CMS:2019leu,CMS:2021ynu,ALICE:2020jff,ALICE:2022cxs,LHCb:2022kph}, have proven essential in testing the universality of the effects that are seen in the hadronic final states. Figure~\ref{fig:jet_ewboson} (right) shows the measurements of $W^+$-boson production from~\cite{CMS:2019leu,ALICE:2022cxs}. The data are presented here as ratios relative to a pQCD baseline that includes isospin effects but no genuine nuclear modifications. Relative to this baseline, the measurements show possible signs of antishadowing enhancement and the EMC effect onset (minding the normalisation uncertainty of the data and the baseline) at negative rapidities of the final state decay lepton, and clear shadowing at positive rapidities. These data are consistent with the observations from the measurements with inclusive hadrons and jets, but might favour slightly less pronounced gluon shadowing.

Measurements on prompt photons~\citep{ATLAS:2019ery,ALICE:2025bnc} also provide a picture on nuclear modifications consistent with other observables, but are currently significantly limited by statistical precision. The same applies for top-quark pair ($t\bar{t}$) production~\citep{CMS:2017hnw,ATLAS:2024qdu}, where the current experimental precision cannot strongly discriminate between predictions with and without nuclear modifications. Since electroweak-boson and $t\bar{t}$ production in leptonic decay channel are insensitive to the QGP formed in the collision, they can be used for studying nuclear PDFs also in $A+A$ collisions. The available data~\citep{ATLAS:2019ibd,ATLAS:2019maq,ATLAS:2024jji,ALICE:2020jff,ALICE:2022cxs,CMS:2021kvd,CMS:2026hhb} are again roughly consistent with the expected nuclear-PDF effects, but would generally benefit from higher statistics.

\subsubsection{Photonuclear processes in UPCs}
\label{subsec:UPCs}

Another important source of information on nuclear PDFs is real photon-nucleus scattering in ultraperipheral collisions (UPCs) of heavy ions at RHIC and the LHC~\citep{Bertulani:2005ru,Baltz:2007kq,Contreras:2015dqa,Klein:2019qfb}. UPCs are characterized by very large transverse distances (impact parameters) $|\mathbf{b}|$ between colliding ions $A$ and $B$, $|\mathbf{b}| \gg R_A+R_B$, where $R_A$ and $R_B$ are the nucleus radii. In this case, the hadronic interactions between nuclei $A$ and $B$ are suppressed, and the interaction proceeds via the emission of quasi-real photons in the equivalent photon approximation, whose flux scales as the nucleus electric charge squared, $N_{\gamma/A} \sim Z^2$, and whose energy scales as $k \sim \gamma_L/R_A$, where $\gamma_L=E_A/m_N$ is the Lorentz factor ($E_A$ is the nuclear beam energy in the collision frame). As a result, UPCs allow one to study photon-nucleus scattering in the collider kinematics of large $W$ and small $x$, which is complementary to that covered by $l^{\pm}+A$ DIS with fixed nuclear target and at the future EIC~\citep{Accardi:2012qut}.

Focusing on hard photonuclear processes, measurements of Pb-Pb UPCs at the LHC include photoproduction of charmonia $J/\psi$~\citep{ALICE:2013wjo,ALICE:2012yye,CMS:2016itn,ALICE:2021gpt,ALICE:2021tyx,ALICE:2023jgu,ALICE:2023gcs,CMS:2023snh,LHCb:2022ahs,ATLAS:2025aav} and $\psi^{\prime}$~\citep{ALICE:2013wjo,ALICE:2021gpt,LHCb:2022ahs}, bottomonia~$\Upsilon$~\citep{CMS:2026soy}, dijets~\citep{ATLAS:2024mvt}, and open charm $D^0$ mesons~\citep{CMS:2025jjx}. In addition, production of $J/\psi$ in Au-Au UPCs~\citep{PHENIX:2009xtn,STAR:2023vvb,STAR:2023nos} and in d-Au UPCs~\citep{STAR:2021wwq} was studied at RHIC. These measurements unambiguously demonstrated modifications of the corresponding nuclear photoproduction cross sections, which are consistent with those encoded in nuclear PDFs. Notably, the data on coherent $J/\psi$ photoproduction, which at the lowest order of pQCD probe the gluon density of the target~\citep{Ryskin:1992ui}, gave the direct evidence of gluon nuclear shadowing in heavy nuclei, $f_g^A(x_N,Q^2)/(Af_g^N(x_N,Q^2))<1$, for a wide interval of small $x_N$, $10^{-5} < x_N < 0.05$, at $Q^2={\cal O}(M^2_{J/\psi})$, where $M_{J/\psi}$ is the mass of the $J/\psi$ vector meson.

\section{Large $x$: Incoherent scatterings off individual nucleons (and non-nucleonic constituents)}
\label{sec:large-x}

For sufficiently large values of $x_N$, a hard interaction with a nuclear target can be seen as incoherent scattering from individual nucleons. Forming a quantum-mechanical system, the nucleons in nuclei are, however, not stationary, and their positions and momenta should be regarded as fluctuating quantities. This quantum motion of nucleons in nuclei, known as \emph{Fermi motion}, causes the nuclear PDFs to be smeared with respect to a distribution built from stationary free nucleons. These smearing effects are typically described in the convolution formalism with
\begin{equation}
  f^A_i(x_N,Q^2) = \sum_{N=p,n} \int_{x_N}^A \, \frac{\mathrm{d}y_N}{y_N} f^{N/A}(y_N) \, f^N_i(x_N/y_N,Q^2).
  \label{eq:conv-form}
\end{equation}
Here, $f^{N/A}(y_N)$ is the light-cone distribution of nucleons carrying a fraction $y_N$ of the average nucleon momentum $P_A/A$, and $f^N_i(x_N/y_N,Q^2)$ are on-shell free-nucleon PDFs for $N \in \{p,n\}$, assuming that there are no other nuclear effects beyond Fermi motion at play. Taking the sum to run over $Z$ protons and $A-Z$ neutrons gives the normalisation $\int_0^A \mathrm{d}y_N f^{N/A}(y_N) = 1$. The distribution $f^{N/A}(y_N)$ is strongly peaked around $y_N = 1$, and thus the Fermi motion impacts the nuclear PDFs only in the large-$x_N$ region. As a result of the smearing, the nuclear PDFs get an enhancement compared to those of free nucleons for large $x_N < 1$, explaining the observed rise in the DIS structure function ratio in this region (cf.\ Fig.~\ref{fig:dis}), and also acquire a high-momentum tail for $x_N > 1$, corresponding to nucleons carrying momentum larger than the average nucleon momentum in the nucleus. A well-accepted conclusion from studies of Fermi-motion is that while it describes the large-$x_N$ enhancement, the quantum motion of otherwise free and unmodified nucleons alone cannot explain the observed EMC effect, and some additional nuclear effects have to be included, see~\cite{Bickerstaff:1989ch}.

A simple explanation of why Fermi motion cannot give the complete picture comes from considering that because nuclei are bound systems, energy conservation constrains the allowed nucleon momentum and removal-energy distributions. Including the binding effects modifies the single-nucleon momentum distribution $f^{N/A}$, leading to a suppression in the EMC region, but at the expense that it no longer satisfies the momentum sum rule alone. One can expand the picture further by discussing how binding is dynamically realised in nuclei. In conventional nuclear models the binding is associated with the exchange of (virtual) mesons, predominantly pions. These particles could therefore be part of the nuclear wave function and contribute to the parton content of nuclei. This would then cause a fraction of the nuclear light-cone momentum to be carried by mesonic degrees of freedom rather than nucleons, and produce an additional pionic contribution to Eq.~\eqref{eq:conv-form}. As a result, the total nuclear valence quark distribution would be shifted to smaller momentum fraction, explaining the EMC suppression. In accounting for such an effect, one should note that a free nucleon can already be viewed to have a pion cloud about it, and the additional contribution should reflect the enhancement of the pion field in nuclei. The increase in the number of pions should also lead to an enhancement in the sea-quark distribution in nuclei, which appears to conflict with the observation of no antishadowing enhancement in the $p+A$ DY data.

The effects discussed above, Fermi motion, binding and pion exchange, are sometimes grouped together under the umbrella of traditional nuclear-physics explanation of the EMC effect. Additional and alternative mechanisms have been also proposed. An important question here is whether the internal structure of the bound nucleons is modified compared to free ones, which seems to be suggested by the above-mentioned challenges of the traditional picture. It has been noted that due to binding the nucleons do not have to be on their mass shell, which can also lead to modifications in the nucleon PDFs $f^N_i$ in Eq.~\eqref{eq:conv-form}. Alternatively, the nucleons can experience either general mean-field modifications due to the presence of the other nucleons in the nucleus, or more localised, e.g.\ clustering or short-range correlation type of fluctuations that originate from nucleon-nucleon interactions and end up altering the average bound-nucleon PDFs. Many of the nuclear effects discussed above do not rule each other out and can even be different manifestations of the same physics. At present, it is not clear what is the dominant mechanism causing the EMC effect, and modern model-dependent determinations of the nuclear modifications can include a combination of multiple different effects that jointly explain the EMC suppression; see, e.g.,~\cite{Kulagin:2004ie}. A more extensive discussion on these issues is available in the respective chapter on the EMC effect in this Encyclopedia. More detailed reviews can also be found in~\cite{Frankfurt:1988nt,Bickerstaff:1989ch,Arneodo:1992wf,Geesaman:1995yd,Norton:2003cb,Malace:2014uea,Hen:2016kwk}.

\section{Small-$x$ shadowing: Multi-nucleon coherence effects}
\label{sec:small-x}

Nuclear shadowing is a general phenomenon of high-energy scattering on nuclear targets, which constitutes the observation that nuclear scattering cross sections are smaller than the sum of cross sections for scattering on individual nucleons. In the case of inclusive $l^{\pm}+A$ DIS, it manifests itself as suppression of the structure function ratio $F_2^A(x,Q^2)/F_2^d(x,Q^2)< 1$ for small $x$, see Sec.~\ref{sec:experiment}. This effect can be explained by coherent multiple scattering of hadronic components of the virtual photon on nucleons of the nuclear target~\citep{Piller:1999wx,Armesto:2006ph}. Note that in this section, we follow the traditional notation for brevity: the momentum fraction $x$ should be identified with $x_N$, and the nuclear structure function ratios are understood to be normalized by the corresponding atomic mass numbers.

In the target rest frame, the incoming high-energy virtual photon $\gamma^{\ast}$ can be viewed as a superposition of long-lived hadronic, e.g.~$\rho, \omega, \phi$ vector mesons~\citep{Bauer:1977iq}, and partonic $q{\bar q}, q {\bar q}g$, etc.~\citep{Nikolaev:1990ja} fluctuations, whose lifetime $l_c$ (also called the coherence length or Ioffe time) can be estimated using the energy-time uncertainty relation,
\begin{equation}
l_c=\frac{1}{2 x m_N} \,,
\label{eq:l_c}
\end{equation}
where $m_N$ is the nucleon mass. For sufficiently small $x$, $x < 0.05$, the lifetime $l_c$ exceeds the average distance between nucleons in nuclei, $r_{NN} \approx 1.7$ fm, and eventually becomes compatible to the nucleus radius $R_A$, $l_c \sim R_A$. As a result, coherent scattering of these fluctuations on several or all target nucleons leads to destructive interference among the amplitudes corresponding to the interaction with different number of nucleons, resulting in suppression of the $\gamma^{\ast} A$ cross section. This can be interpreted as being caused by nucleons eclipsing each other so that nucleons in the back of the nucleus lie in the geometric shadow of the ones in the front~\citep{Glauber:1955qq}.

\subsection{Gribov-Glauber theory of nuclear shadowing}

\begin{figure}[t]
\centering
\includegraphics[height=4cm]{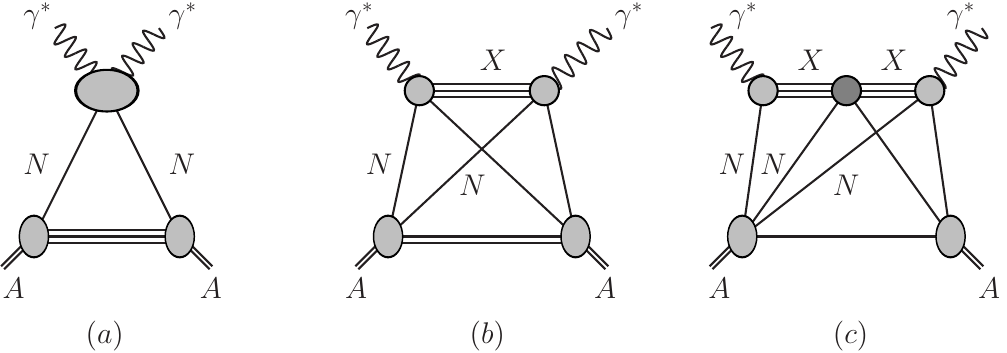}
\caption{The Gribov-Glauber series for the forward $\gamma^{\ast} A$ amplitude in terms of interaction with one (a), two (b), and three (c) target nucleons, where graph (a) represents the impulse approximation, and graphs (b) and (c) contribute to the shadowing correction. The ovals denote the interaction and nuclear vertices, and $X$ stand for the intermediate states produced in $\gamma^{\ast}N$ interactions.}
\label{fig:GG_2026}
\end{figure}

This space-time picture of high-energy $\gamma^{\ast}+A$ scattering leads to the Gribov-Glauber theory of nuclear shadowing~\citep{Gribov:1968jf,Glauber:1970jm}. In this approach, the total  virtual photon-nucleus ($\gamma^{\ast} A$) cross section, $\sigma_{\gamma^{\ast} A}$, can be presented as a series of terms corresponding to the interaction with $N=1, \dots, A$ target nucleons,
\begin{equation}
 \sigma_{\gamma^{\ast} A}(x,Q^2)=\sum_{N=1}^{A} \sigma_{\gamma^{\ast} A}^{(N)}(x,Q^2) =A  \sigma_{\gamma^{\ast} N}(x,Q^2)-\delta \sigma_{\gamma^{\ast} A}(x,Q^2) \,,
 \label{eq:GG}
\end{equation}
where $\sigma_{\gamma^{\ast} N}$ is the total virtual photon-nucleon ($\gamma^{{\ast}} N$) cross section. Thus, the $N=1$ term represents the impulse approximation (here we safely neglected quantum motion of nucleons in nuclei discussed in Sec.~\ref{sec:large-x}), and the $N \geq 2$ terms build up the shadowing correction $\delta \sigma_{\gamma^{\ast} A}$. This is illustrated in Fig.~\ref{fig:GG_2026}, which shows the contributions to the forward $\gamma^{\ast}A$ amplitude from the interaction with one ($N=1)$, two ($N=2$), and three ($N=3$) nucleons of the target nucleus. (We do not show higher terms with $N \geq 4$.) The ovals denote the interaction and nuclear vertices, and $X$ stand for the intermediate states produced in  $\gamma^{\ast}N$ interactions.

In the limit of low nuclear thickness, i.e.\ for light-to-intermediate nuclei, or for not too small $x$, the dominant contribution to nuclear shadowing comes from the double scattering  graph (b), which gives
\begin{equation}
\delta \sigma_{\gamma^{\ast} A}^{(b)}(x,Q^2) =8 \pi \int d^2  {\bf b} \int_{-\infty}^{\infty} dz_1 \, \rho_A({\bf b},z_1) \int_{z_1}^{\infty} dz_2 \, \rho_A({\bf b},z_2) \int_{4 m_{\pi}^2}^{W^2} dM_X^2 \cos((z_1-z_2)/\lambda)
\frac{d^2\sigma^{\rm diff}_{\gamma^{\ast} N}}{dt dM_X^2}\Bigg|_{t \approx 0} \,,
\label{eq:delta_sigma_2}
\end{equation}
where ${\mathbf b}$ and $z_{1,2}$ are the transverse and longitudinal coordinates of the two nucleons involved in the interaction with the virtual photon, $\rho_A$ is the nuclear density (we used the independent nucleon approximation neglecting short-range nucleon-nucleon correlations), and $d^2\sigma^{\rm diff}_{\gamma^{\ast} N}/(dt dM_X^2)$ is the $\gamma^{\ast}N \to XN$ diffractive cross section with $t$ the momentum transfer squared. The integration over the diffractive masses $M_X$ corresponds to summing the intermediate states $X$ in Fig.~\ref{fig:GG_2026}; it runs from the lower limit, determined by production of two pions with the mass $m_{\pi}$, to the upper limit given by the total center-of-mass $\gamma^{\ast}N$ energy $W$. Finally, the parameter $\lambda=1/[x m_N (1+M_X^2/Q^2)]$ plays the role of the coherence length $l_c$, see Eq.~(\ref{eq:l_c}), so that the damping factor $\cos((z_1-z_2)/\lambda)$, which accounts for a finite longitudinal momentum transfer in the $\gamma^{\ast}N \to X N$ process~\citep{Bertocchi:1972cj}, suppresses the contribution of large diffractive masses $M_X$, whose associated lifetime $\lambda$ exceeds the average inter-nucleon distance $r_{NN}$. Note that since the $t$ dependence of the $\gamma^{\ast}N$ diffractive cross section is much weaker than that of the nuclear form factor calculated through the nuclear density $\rho_A$, the diffractive cross section is evaluated at $t\approx 0$, placing the nucleons at the same transverse distance ${\mathbf b}$.

The key feature of Eq.~(\ref{eq:delta_sigma_2}) is that it allows one to express the shadowing correction to the total $\gamma^{\ast}A$ cross section in terms of the $\gamma^{\ast}N \to XN$ diffractive cross section. It prescribes  that the intermediate state $X$ in Fig.~\ref{fig:GG_2026} should contain all possible diffractively produced elastic~\citep{Glauber:1955qq} and inelastic~\citep{Gribov:1968jf} states (the latter is referred to as Gribov inelastic shadowing). While the relation between nuclear shadowing and diffraction appears intuitively straightforward, its field-theoretic derivation (including the overall sign) is rather non-trivial and uses so-called Abramovski-Gribov-Kancheli unitarity cutting rules~\citep{Abramovsky:1973fm,Bertocchi:1976bq}. Note that for DIS on deuterium, graph (b) gives the complete expression for the shadowing correction, which allows one to test the shadowing-diffraction connection against available data on the deuteron structure function $F_2^d$~\citep{Melnitchouk:1992eu,Piller:1997ny}.

While the contribution of the double scattering term ($N=2$) to nuclear shadowing and its connection to diffraction are well-established, accounting for higher terms with $N\geq 3$ in Eq.~(\ref{eq:GG})  for intermediate and heavy nuclei requires certain modeling~\citep{Kopeliovich:1995ju,Capella:1997yv,Melnitchouk:1993vc,Piller:1995kh,Armesto:2003fi,Armesto:2010kr}. It is customary to assume that these contributions can be expressed in terms of the effective cross section $\sigma_{\rm eff}$, which parametrizes the state $X$-nucleon interaction (the central dark circle in graph (c) of Fig.~\ref{fig:GG_2026}),
\begin{equation}
\sigma_{\rm eff}=\frac{16 \pi}{\sigma^{\gamma^{\ast}N}(x,Q^2)} \int_{4 m_{\pi}^2}^{W^2} dM_X^2
\frac{d^2\sigma^{\rm diff}_{\gamma^{\ast} N}}{dt dM_X^2}\Bigg|_{t \approx 0} \,.
 \label{eq:sigma_eff}
\end{equation}
Summing all the contributions with help of the eikonal approximation, one obtains the final expression for the shadowing correction,
\begin{equation}
\delta \sigma_{\gamma^{\ast} A}(x,Q^2) =8 \pi \int d^2  {\bf b} \int_{-\infty}^{\infty} dz_1 \, \rho_A({\bf b},z_1) \int_{z_1}^{\infty} dz_2 \, \rho_A({\bf b},z_2) \int_{4 m_{\pi}^2}^{W^2} dM_X^2 \cos((z_1-z_2)/\lambda)
\frac{d^2\sigma^{\rm diff}_{\gamma^{\ast} N}}{dt dM_X^2}\Bigg|_{t \approx 0}
\exp\left[-\frac{\sigma_{\rm eff}}{2} \int_{z_1}^{z_2} dz^{\prime} \rho_A({\bf b},z^{\prime})\right]\,.
\label{eq:delta_sigma_all}
\end{equation}

Application of Eq.~(\ref{eq:delta_sigma_all}) for the calculation of nuclear shadowing in $l^{\pm}+A$ DIS requires as input the nucleon (proton) diffractive cross section $d^2\sigma^{\rm diff}_{\gamma^{\ast} N}/(dt dM_X^2)$. Using the two-component hadronic structure of the virtual photon discussed in the beginning of this section, this cross section can be written in the form of a mass dispersion relation, where the low-mass contribution is approximated by a sum of $\rho, \omega, \phi$ vector mesons and the high-mass contribution is expressed through the proton diffractive structure function $F_2^{D(4)}$,
\begin{equation}
\frac{d^2\sigma^{\rm diff}_{\gamma^{\ast} N}}{dt dM_X^2}=\sum_{V=\rho, \omega, \phi}
\frac{4 \pi \alpha_{\rm em}}{f_{V}^2} \frac{M_V^4}{(M_V^2+Q^2)^2} \frac{\sigma_V^2}{16 \pi} \delta(M_X^2-M_V^2)+\frac{4 \pi^2 \alpha_{\rm em}}{Q^2} F_2^{D(4)}(x,Q^2;x_{\Pomeron},t)\left|\frac{dx_{\Pomeron}}{dM_X^2} \right| \theta(M_X^2-M_{X,0}^2)\,.
\label{eq:sigma_diff}
\end{equation}
In the first term, the vector meson dominance (VMD) model is used in the limit of infinitely narrow vector mesons, where  $\alpha_{\rm em}$ is the fine-structure constant, $f_V$ is the photon-meson coupling constant determined through the $\Gamma(V \to e^{+}e^{-})$ leptonic decay width, $M_V$ is the vector meson mass, and $\sigma_V$ is the vector meson-nucleon cross section. The second term involves the structure function $F_2^{D(4)}$, which has been measured in inclusive diffraction in $l^{\pm}+p$ DIS at Hadron-Electron Ring Accelerator (HERA); for reviews, see~\cite{Klein:2008di,Newman:2013ada,Frankfurt:2022jns}. In addition to the Bjorken variables $x$ and $Q^2$, it depends on the momentum fraction $x_{\Pomeron}=(Q^2+M_X^2)/(W^2+Q^2)$, which can be interpreted as the Pomeron longitudinal momentum fraction, and $t$. Note that we also included the Jacobian $|dx_{\Pomeron}/dM_X^2|=x/Q^2$ for the $M_X^2 \to x_{\Pomeron}$ variable transformation and the Heaviside step function $\theta(M_X^2-M_{X,o}^2)$ to indicate the minimal mass $M_{X,0} \geq 2$ GeV produced in diffractive DIS.

Substituting Eq.~(\ref{eq:sigma_diff}) in Eq.~(\ref{eq:delta_sigma_all}), the latter can be readily turned into an expression for the shadowing correction to the nuclear structure function, $\delta F_2=A F_2^N-F_2^A$,
\begin{multline}
\delta F_2^A(x,Q^2) =
\frac{Q^2}{\pi} \int d^2  {\bf b} \int_{-\infty}^{\infty} dz_1 \, \rho_A({\bf b},z_1) \int_{z_1}^{\infty} dz_2 \, \rho_A({\bf b},z_2)
\sum_{V=\rho, \omega, \phi} \frac{M_V^4}{f_V^2(M_V^2+Q^2)^2}
\cos((z_1-z_2)/\lambda_V)
\frac{\sigma_V^2}{2}
\exp\left[-\frac{\sigma_{V}}{2} \int_{z_1}^{z_2} dz^{\prime} \rho_A({\bf b},z^{\prime})\right] \\
 +
8 \pi \int d^2  {\bf b} \int_{-\infty}^{\infty} dz_1 \, \rho_A({\bf b},z_1) \int_{z_1}^{\infty} dz_2 \, \rho_A({\bf b},z_2) \int_{x}^{0.1} dx_{\Pomeron} \cos((z_1-z_2)/\lambda) F_2^{D(4)}(x,Q^2; x_{\Pomeron}, t \approx 0)
\exp\left[-\frac{\tilde{\sigma}_{\rm eff}}{2} \int_{z_1}^{z_2} dz^{\prime} \rho_A({\bf b},z^{\prime})\right]\,,
\label{eq:delta_F_2A}
\end{multline}
where $\lambda_V=1/[x m_N (1+M_V^2/Q^2)]$ and $\tilde{\sigma}_{\rm eff}$ is calculated using Eq.~(\ref{eq:sigma_eff}), where one now retains only the contribution of large masses $M_X \geq M_{X,0}$. The limits of integration over $x_{\Pomeron}$ are determined by the kinematic constraint $x_{\Pomeron} \geq x$ and the standard condition on the produced diffractive masses $M_X^2/W^2 \leq 0.1$. Equation~(\ref{eq:delta_F_2A}) explicitly demonstrates that the two terms in Eq.~(\ref{eq:sigma_diff}) provide complementary contributions to nuclear shadowing: while the VMD term is important at low $Q^2$, giving only a small, higher-twist effect for $Q^2 > 1$ GeV$^2$, the second term proportional to the leading twist $F_2^{D(4)}$ dominates in the scaling limit of large $Q^2$.

\begin{figure}[t]
\centering
\includegraphics[height=6.5cm]{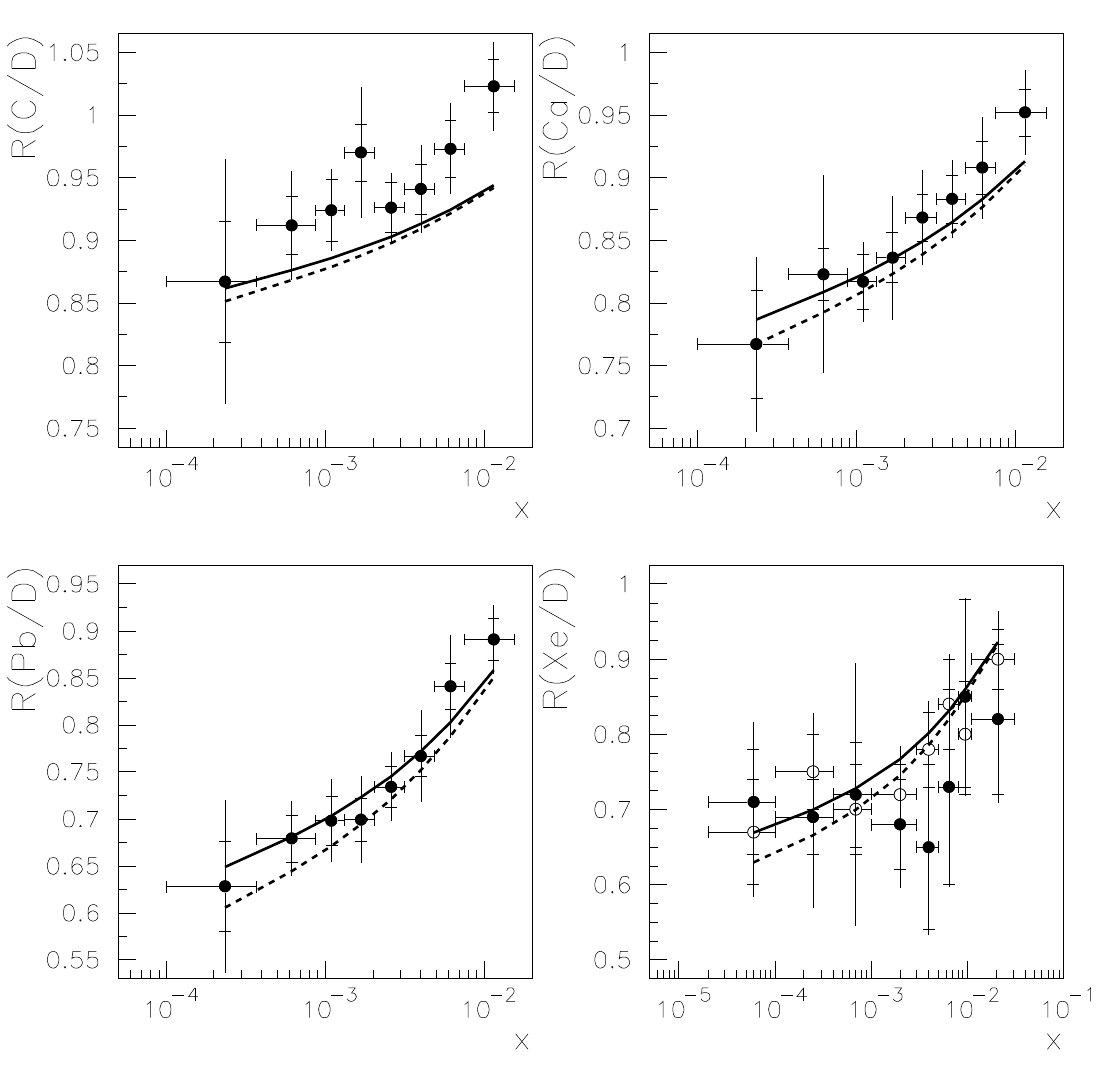}
\includegraphics[height=6.5cm]{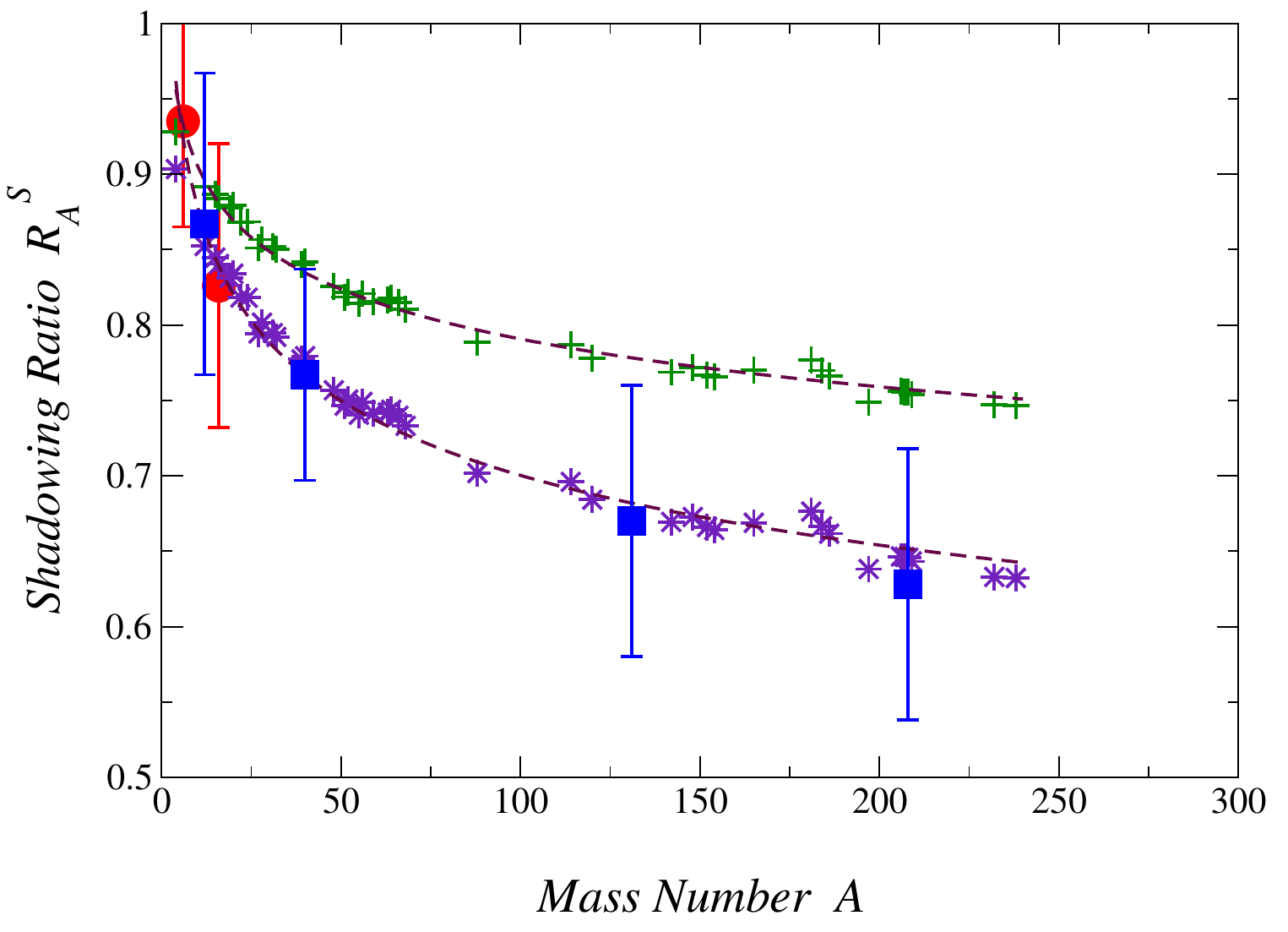}
\caption{Gribov-Glauber theory predictions and comparison with nuclear fixed-target data. (Left) Calculations for $R=F_2^A/F_2^d$~\citep{Armesto:2003fi} vs.~Fermilab E665 data~\citep{E665:1995xur,E665:1992ivw}. The solid and dotted curves correspond to Schwimmer and eikonal approximations. Reprinted from~\cite{Armesto:2003fi}, with permission from Springer Nature. (Right) Calculations for $R_A^S=\sigma_{\gamma A}/\sigma_{\gamma N}$ \citep{Adeluyi:2006xy} at $W=15$ GeV (green crosses) and $W=25$ GeV (purple stars) vs.~CERN NMC~\citep{NewMuon:1995cua,NewMuon:1995tgs} (red circles) and Fermilab E665~\citep{E665:1995xur,E665:1992ivw} (blue squares) data. Reprinted figure with permission from~\cite{Adeluyi:2006xy}. Copyright (2006) by the American Physical Society.}
\label{fig:shadowing_Capella}
\end{figure}

One finds that the resulting Gribov-Glauber theory of nuclear shadowing describes well the ratios of the nuclear and deuteron structure functions $F_2^A(x,Q^2)/F_2^d(x,Q^2)$ for small $x$ and in a wide range of $Q^2$ and nuclear targets~\citep{Piller:1999wx,Armesto:2006ph}, see Fig.~\ref{fig:shadowing_Capella} (left) for an example. Note that this approach to nuclear shadowing is also valid in the photoproduction ($Q^2=0$) limit. In particular, using the Fermilab data on diffraction dissociation of real photons on hydrogen~\citep{Chapin:1985mf}, one can successfully describe experimental data on the ratio of the nuclear and nucleon photoabsorption cross sections, $\sigma_{\gamma A}/\sigma_{\gamma N}$, for a large array of nuclei~\citep{Adeluyi:2006xy}, see the right panel of Fig.~\ref{fig:shadowing_Capella}.

\subsection{Leading twist approximation (LTA) for nuclear shadowing}

The Gribov-Glauber theory for nuclear structure functions $F_2^A$ can be extended to individual (i.e.~separately for quarks and gluons) nuclear PDFs $f_{i}^A$. It is made possible by the QCD factorization theorem for diffractive hard scattering~\citep{Collins:1997sr}, allowing for QCD analyses of inclusive diffraction in $l^{\pm}+p$ DIS at HERA~\citep{H1:2006uea,H1:2006zyl,ZEUS:2009uxs}. The resulting framework is referred to in the literature as the leading twist approximation (LTA) for nuclear shadowing~\citep{Frankfurt:1998ym,Frankfurt:2011cs} because it relies on the experimental observation that the diffractive structure function $x_{\Pomeron}F_2^{D(3)}$ (reduced cross section $x_{\Pomeron}\sigma_r^{D(3)}$) is approximately independent of the virtuality $Q^2$ in a wide kinematic  window of $3.5 < Q^2< 1600$ GeV$^2$, $0.0003 < x_{\Pomeron} < 0.03$ and $0.0017 < x/x_{\Pomeron} < 0.8$, suggesting its leading twist partonic interpretation.

Starting with the second line of Eq.~(\ref{eq:delta_F_2A}) and applying the factorization theorems for inclusive $l^{\pm}+A$ DIS~\citep{CTEQ:1993hwr} to the left-hand side and for diffraction in $l^{\pm}+p$ DIS to the right-hand side, one obtains the following expression for the shadowing correction to nuclear PDFs $\delta f_{i}^A=Af_i^N-f_i^A$,
\begin{equation}
\begin{split}
 \delta f_i^A(x,Q^2) & =
8 \pi \, \Re e  \frac{(1-i \eta)^2}{1+\eta^2}\int d^2  {\bf b} \int_{-\infty}^{\infty} dz_1 \, \rho_A({\bf b},z_1) \int_{z_1}^{\infty} dz_2 \, \rho_A({\bf b},z_2) \int_{x}^{0.1} \frac{dx_{\Pomeron}}{x_{\Pomeron}} \exp[i(z_1-z_2)x_{\Pomeron}m_N ] f_i^{D(4)}(x,Q^2; x_{\Pomeron}, t \approx 0) \\
&\quad\times
\exp\left[-\frac{(1-i \eta)\sigma_{\rm eff}^i}{2} \int_{z_1}^{z_2} dz^{\prime} \rho_A({\bf b},z^{\prime})\right]\,,
\label{eq:delta_f}
\end{split}
\end{equation}
where $f_i^{D(4)}$ are proton diffractive PDFs~\citep{,H1:2006uea,H1:2006zyl,ZEUS:2009uxs,Goharipour:2018yov,Salajegheh:2022vyv,Salajegheh:2023jgi}, which are conditional probabilities of finding partons in the proton, provided that it elastically scatters into the final system carrying the longitudinal momentum fraction $1-x_{\Pomeron}$ and the momentum transfer $t$; $\eta=\pi/2(\alpha_{\Pomeron}(0)-1)\approx 0.17$ is a small correction for the non-zero real part of the $\gamma^{\ast}N \to XN$ amplitude, which is estimated using the dispersion relation and the Pomeron intercept $\alpha_{\Pomeron}(0)$ from the fits to HERA data~\citep{H1:2006zyl}.

The main theoretical uncertainty of the LTA approach stems from modeling the effective parton cross section $\sigma^i_{\rm eff}$. Intuitively, it is related to the probability of diffraction in a given partonic channel, which can be quantified in terms of the cross section $\sigma_2^i$ [c.f.~Eq.~(\ref{eq:sigma_eff})]
\begin{equation}
\sigma_{2}^i=\frac{16 \pi}{(1+\eta^2)f_i^N(x,Q^2)} \int_{x}^{0.1}  \frac{dx_{\Pomeron}}{x_{\Pomeron}}
f_i^{D(4)}(x,Q^2; x_{\Pomeron}, t \approx 0)
\,,
\label{eq:sigma_2}
\end{equation}
where $f_i^N$ are usual proton PDFs and the subscript ``2'' makes a reference to graph (b) in Fig.~\ref{fig:GG_2026}. However, graphs (c) and higher in Fig.~\ref{fig:GG_2026} cannot be unambiguously expressed in terms of $\sigma_2^i$, which leads only to the constraint $\sigma^i_{\rm eff} \geq \sigma_2^i$. In practice, one varies $\sigma^i_{\rm eff}$ between its upper limit given by the pion-nucleon cross section (in the spirit of the VMD model) and its lower limit defined by $\sigma_2^i$ or the $q{\bar q}$ dipole-nucleon cross section (whichever is larger); the corresponding predictions are referred to as the ``low shadowing'' and ``high shadowing'' scenarios.

\begin{figure}[t]
\centering
\includegraphics[height=6.cm]{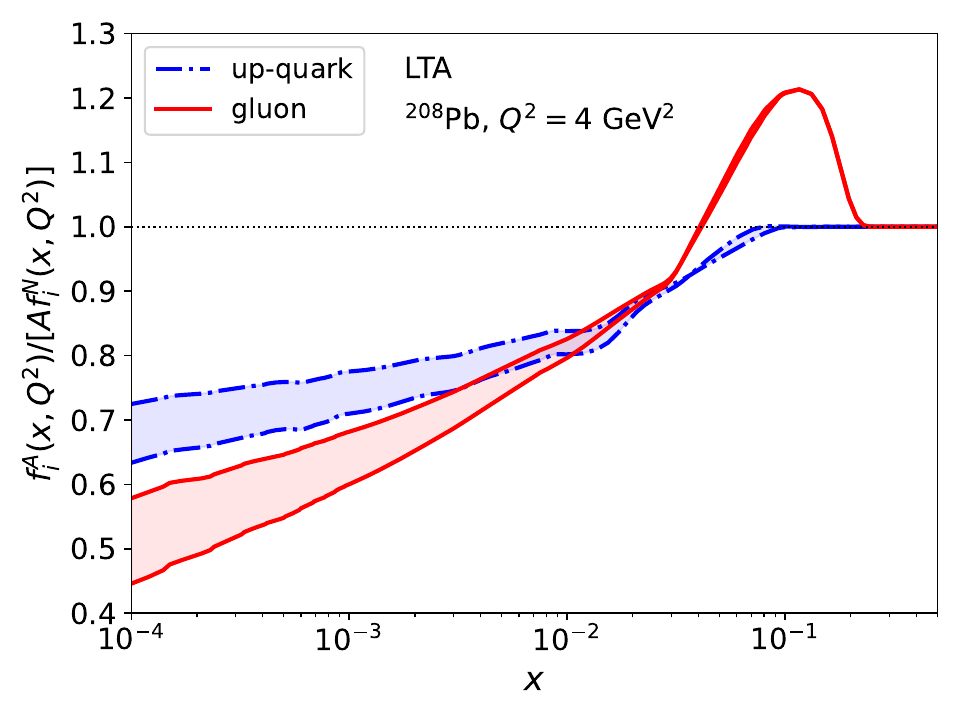}
\includegraphics[height=6.cm]{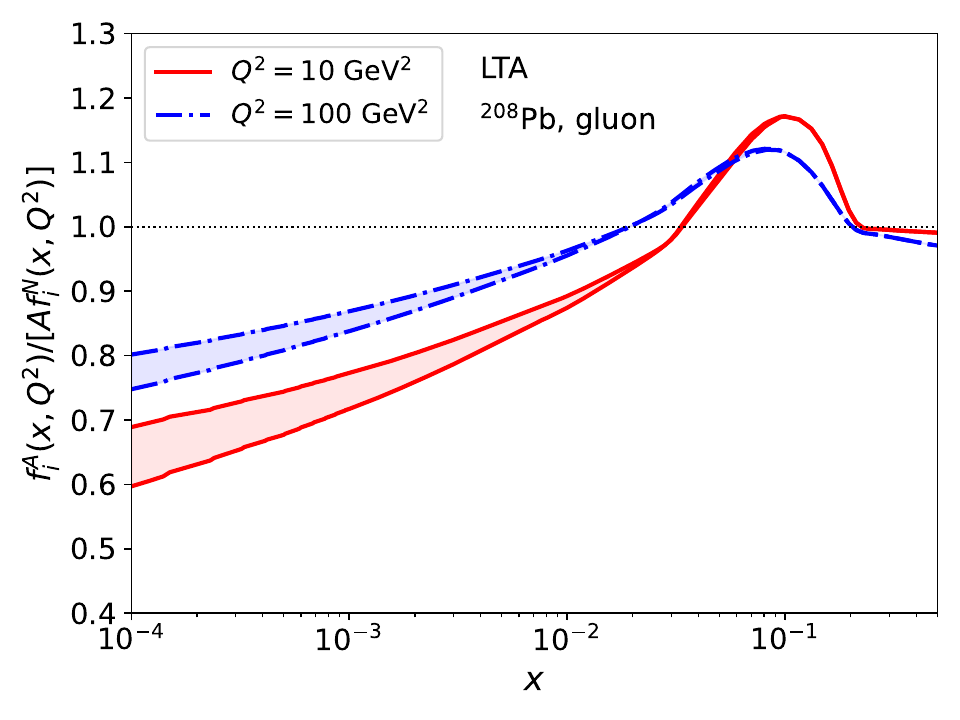}
\caption{(Left) LTA predictions for $f_i^A/(Af_i^N)$ for the up-quark and gluon distributions in $^{208}$Pb as a function of $x$ at $Q^2=4$ GeV$^2$. The uncertainty bands are bounded by ``low shadowing'' and ``high shadowing'' scenarios. (Right) The $Q^2$ dependence of $f_i^A/(Af_i^N)$ for the gluon distribution in $^{208}$Pb.}
\label{fig:LTA_nPDFs_input}
\end{figure}

The characteristic property of LTA is that the experimental observation of significant leading twist diffraction in $l^{\pm}+p$ DIS at HERA translates into relatively sizable diffractive PDFs of the proton, corresponding to large $\sigma_2^i$ and $\sigma_{\rm eff}^i$, which results in large leading twist nuclear shadowing, especially for the gluon distribution. Figure~\ref{fig:LTA_nPDFs_input} (left) shows the LTA predictions for $f_i^A/(Af_i^N)=1-\delta f_i^A/(Af_i^N)$ for the up-quark and gluon distributions in $^{208}$Pb as a function of $x$ at $Q^2=4$ GeV$^2$. The presented results are calculated using Eq.~(\ref{eq:delta_f}) with the H1 2006 DPDF Fit B for $f_i^{D(4)}$~\citep{H1:2006zyl} and the CTEQ5M parametrization for $f_i^N$~\citep{Lai:1999wy}. The shaded bands represent the theoretical uncertainties associated with the variation of $\sigma_{\rm eff}^i$ between the ``low shadowing'' and ``high shadowing'' limits. Note that to compensate for depletion due to nuclear shadowing at low $x$, an enhancement of the nuclear gluon distribution around $x \approx 0.1$ (antishadowing) was introduced by requiring conservation of the momentum sum rule for nuclear PDFs. One can see from the figure that LTA predicts a strong suppression of $f_i^A/(Af_i^N)$ for $x < 0.05$, which is larger for gluons than for quarks and which increases with a decrease of $x$. Important evidence in favor of significant gluon nuclear shadowing has been obtained from analyses of coherent $J/\psi$ photoproduction in Pb-Pb UPCs at the LHC~\citep{Guzey:2013xba,Guzey:2013qza,Guzey:2024gff}, see Sec.~\ref{subsec:UPCs}.

Equation~(\ref{eq:delta_f}) should be applied to calculate nuclear PDFs at some initial scale $Q_0^2$ ($Q_0^2=4$ GeV$^2$ in the considered example), which cannot be too low to avoid potential higher-twist effects enhanced in diffraction~\citep{Motyka:2012ty,Maktoubian:2019ppi,Salajegheh:2022vyv}. Using it as a boundary condition, the subsequent $Q^2$ dependence of nuclear PDFs is given by the DGLAP equations, Eq.~(\ref{eq:dglap}). The right panel of Fig.~\ref{fig:LTA_nPDFs_input} shows $f_i^A/(Af_i^N)$ for the gluon distribution in $^{208}$Pb evolved up to $Q^2=10$ and 100 GeV$^2$. One can see from the figure that the suppression due to nuclear shadowing and the enhancement due to antishadowing as well as the uncertainty bands gradually (logarithmically) decrease with an increase of $Q^2$.

\subsection{Nuclear shadowing in dipole model}

The color dipole model~\citep{Nikolaev:1990ja,Mueller:1993rr} can be viewed as a particular realization of the space-time picture of $\gamma^{\ast}+A$ scattering discussed in the beginning of this section. One assumes that the virtual photon fluctuates into a superposition of quark-antiquark pairs (dipoles) of fixed transverse sizes $r=|\mathbf{r}|$, which elastically scatter on nucleons of a nuclear target. Thus, in the eikonal approximation, the total nuclear photoabsoprtion cross section can be written in the following intuitively transparent form~\citep{Armesto:2006ph},
\begin{equation}
\sigma_{\gamma^{\ast}A}(x,Q^2)=2 \int d^2 {\bf r}\, \rho(r,Q^2) \int d^2  {\bf b} \left(1-\exp\left[-\frac{\sigma_{q {\bar q}}(x,r,\mathbf{b})}{2}T_A(\mathbf{b})\right]\right) \,,
\label{eq:sigma_A_dipole}
\end{equation}
where $\rho(r,Q^2)$ is the probability of the $\gamma^{\ast} \to q {\bar q}$ transition expressed through the photon light-cone wave function $\Psi_{\gamma^{\ast}}$, $T_A(\mathbf{b})=\int dz \rho_A(\mathbf{b},z)$ is the nuclear thickness function, and $\sigma_{q {\bar q}}$ is the dipole-nucleon cross section. The multiple scattering mechanism encoded in Eq.~(\ref{eq:sigma_A_dipole}) does not explicitly include inelastic intermediate states and, hence, does not rely on the connection between nuclear shadowing and diffraction. Also, it assumes a very high energy limit, allowing one to neglect the effects of a finite coherence length.

\begin{figure}[t]
\centering
\includegraphics[height=7.2cm]{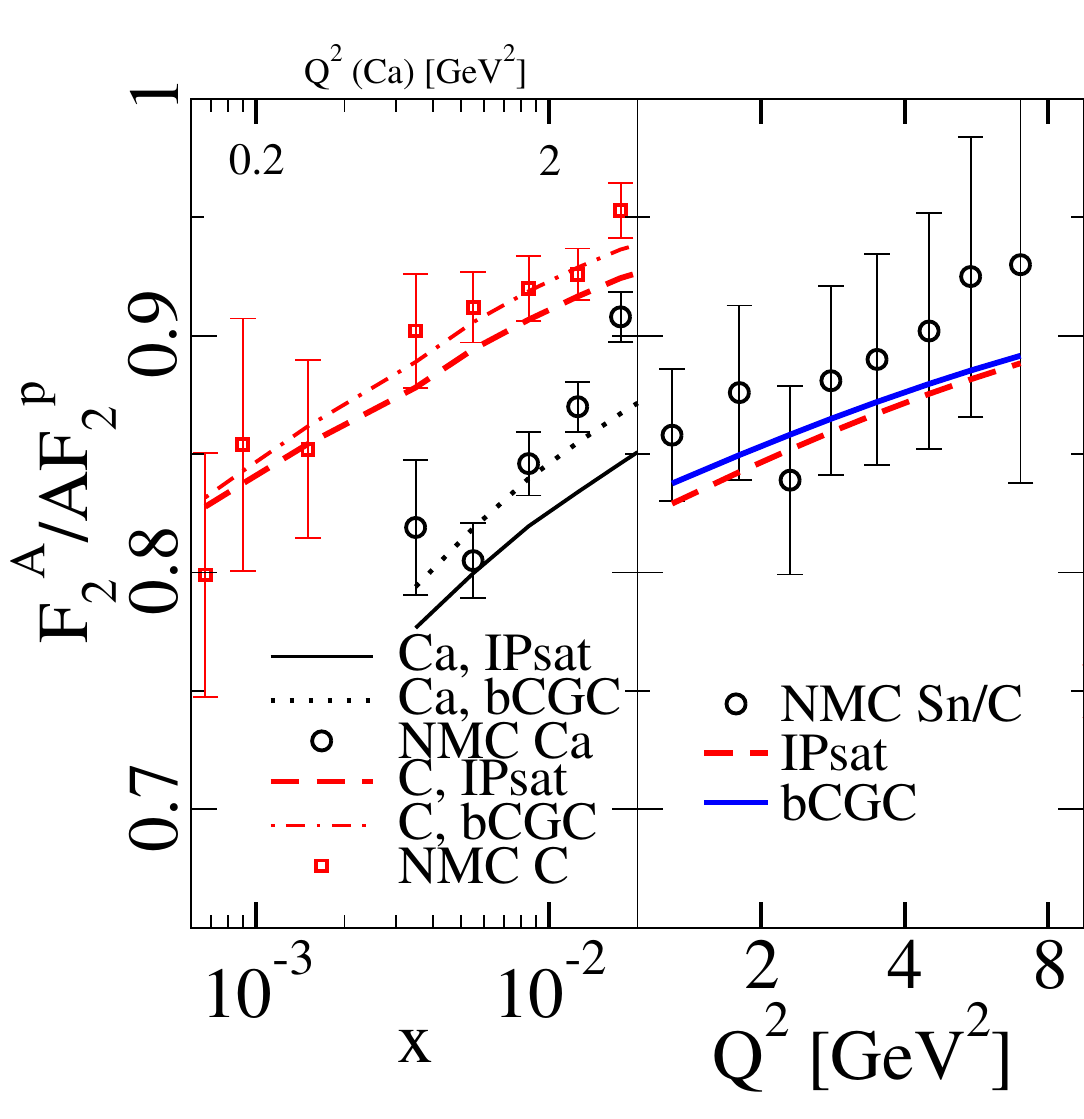}
\includegraphics[height=7.5cm]{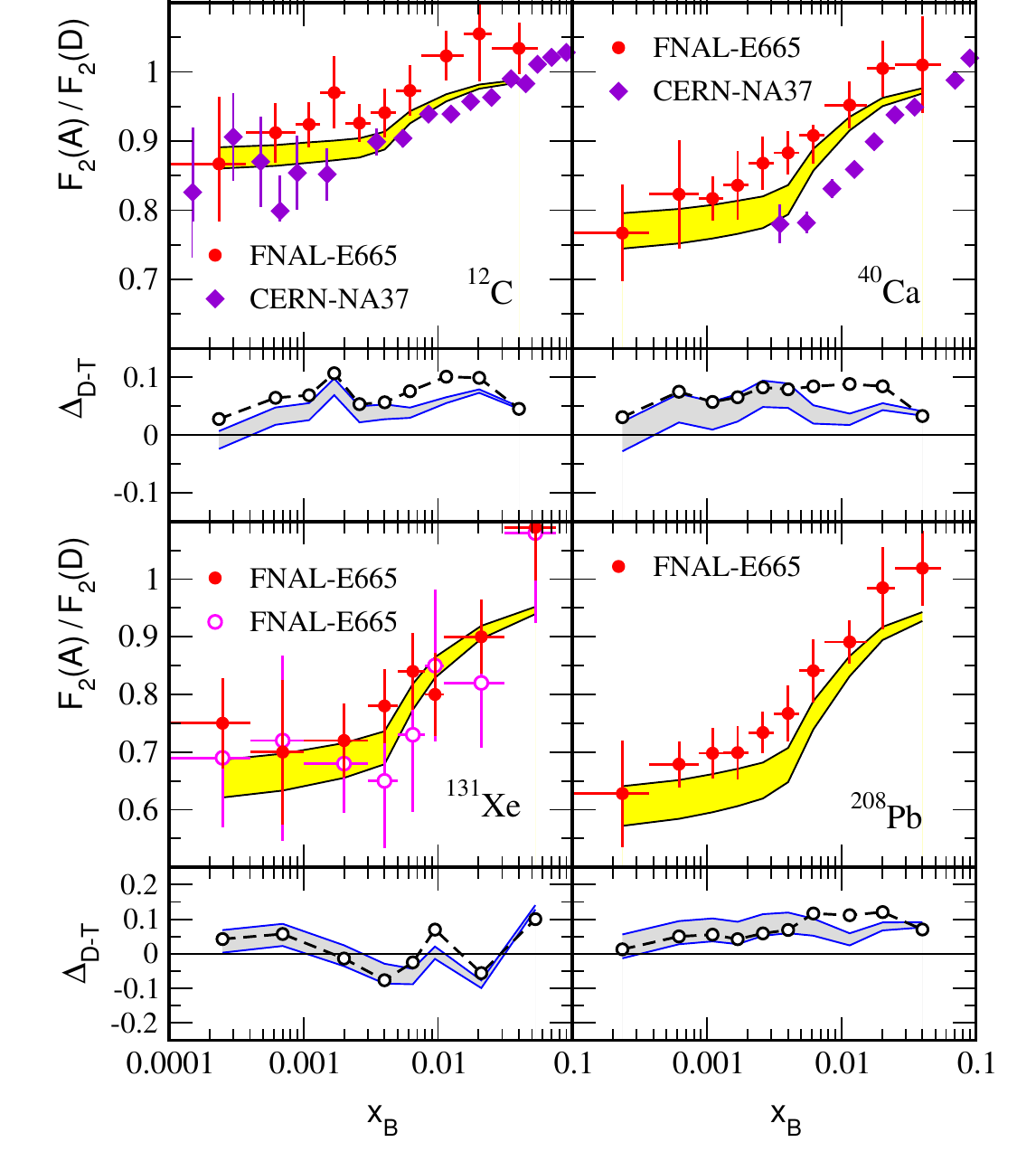}
\caption{(Left) Dipole model predictions for $F_2^A/F_2^p$ and $F_2^A/F_2^C$~\citep{Kowalski:2007rw} using different models for the dipole cross section vs.~CERN NMC data~\citep{NewMuon:1995cua}. Reprinted figure with permission from~\cite{Kowalski:2007rw}. Copyright (2008) by the American Physical Society. (Right) Calculations for $F_2^A/F_2^d$ within resummation of higher twist corrections~\citep{Qiu:2003vd} vs.~Fermilab E665~\citep{E665:1995xur,E665:1992ivw} and CERN NMC~\citep{NewMuon:1995cua} data. The yellow bands correspond to the variation of $\xi^2$. The data minus theory differences, $\Delta_{D-T}$, show also comparison with the EKS98 parametrization~\citep{Eskola:1998df} (gray shaded bands). Reprinted figure with permission from~\cite{Qiu:2003vd}. Copyright (2004) by the American Physical Society.}
\label{fig:shadowing_dipole}
\end{figure}

In the dipole model, the effect of nuclear shadowing originates from nuclear absorption of $q{\bar q}$ dipoles, whose magnitude is controlled by the cross section $\sigma_{q {\bar q}}$ containing details of small-$x$ QCD dynamics. It is usually fitted to HERA data on the proton structure function $F_2^p$ and includes saturation effects, which limit the growth of $\sigma_{q {\bar q}}$ for large $r$ and small $x$~\citep{Golec-Biernat:1998zce,Kowalski:2003hm,Iancu:2003ge,Kowalski:2006hc}. Numerically, $\sigma_{q {\bar q}}$ saturates around $\sigma_0 = 23-29$ mb,  which is large and compatible with typical values of the vector meson-nucleon cross section $\sigma_V$ used in Eq.~(\ref{eq:delta_sigma_all}). As a result, applications of the dipole model to DIS with fixed nuclear targets provide a reasonable description of data on the ratio of nuclear structure functions ($F_2^A/F_2^d$ and $F_2^A/F_2^C$) for small $x$ and small-to-moderate values of $Q^2$~\citep{Armesto:2002ny,Kowalski:2007rw}, which is illustrated in Fig.~\ref{fig:shadowing_dipole} (left).

Using the Green's function technique, the eikonal approximation for the $\sigma_{\gamma^{\ast}A}$ cross section can be generalized to include also higher Fock components of the virtual photon, notably the quark-antiquark-gluon $q {\bar q}g$ fluctuations, as well as the effect of a finite coherence length~\citep{Kopeliovich:2000ra,Kopeliovich:2008ek}. This allows one to effectively take into account Gribov inelastic shadowing, which brings the dipole model conceptually closer to the LTA approach, with a characteristic prediction of large leading twist gluon nuclear shadowing. The latter leads to an improved (compared to keeping only $q{\bar q}$ dipoles~\citep{Lappi:2013am,Mantysaari:2017dwh,Goncalves:2017wgg,Luszczak:2019vdc}) description of coherent $J/\psi$ photoproduction in Pb-Pb UPCs at the LHC~\citep{Kopeliovich:2020has,Luszczak:2024kgi}.

In summary, within all of the approaches considered so far -- the Gribov-Glauber theory, the LTA, and the color dipole model -- nuclear shadowing is explained by coherent multiple scattering on target nucleons. Differences come primarily from various approximations for the hadronic components of the virtual photon and modeling of their interactions with nucleons, which makes it challenging to distinguish nuclear shadowing from parton saturation.

In contrast, treating lepton-nucleus DIS at small $x$ within the framework of Color Glass Condensate (CGC) of high-density QCD~\citep{Iancu:2003xm,Gelis:2010nm}, one can circumvent the notion of a dipole-nucleon cross section. In this approach, the $\sigma_{\gamma^{\ast}A}$ cross section can be expressed in terms of the nuclear dipole scattering amplitude ${\cal N}_A$ as follows~\citep{Armesto:2002ny,Marquet:2017bga,Cepila:2020xol}
\begin{equation}
\sigma_{\gamma^{\ast}A}(x,Q^2)=2 \int d^2 {\bf r}\, \rho(r,Q^2) \int d^2  {\bf b}
{\cal N}_A(x,r,b)=
2 \int d^2 {\bf r}\, \rho(r,Q^2) \int d^2  {\bf b} \left(1-\exp\left[-\frac{Q_{s,A}^2(x,\mathbf{b}) r^2}{4}\right]\right) \,,
\label{eq:sigma_A_BK}
\end{equation}
where $Q_{s,A}^2$ is the nuclear saturation scale, the typical momentum scale below which the gluon density saturates in the nuclear wave function. In CGC phenomenology, defining the saturation scale as the gluon density per unit area, $Q_s^2 \sim \alpha_s xg/(\pi R^2)$, one often writes $Q_{s,A}^2(x,\mathbf{b})=(\sigma_0/2) T_A(\mathbf{b}) Q_{s,p}^2(x)$, where $\sigma_0/2=\pi R_p^2$ is the geometic dipole cross section and $R_p$ is the proton radius. As a result, the effect of nuclear shadowing originates from the parametric nuclear enhancement of $Q_{s,A}^2 \sim A^{1/3}$ with respect to $Q_{s,p}^2$ of the proton. While one can in principle obtain a reasonable description of fixed-target data on $F_2^A/F_2^p$ by treating $Q_{s,A}^2$ as a fit parameter~\citep{Freund:2002ux,Cepila:2020xol}, one also observed certain tensions~\citep{Armesto:2002ny}. Note that this approach provides a competetively good descrition of coherent and incoherent $J/\psi$ photoproduction in heavy ion UPCs at the LHC, including the $t$ dependence of the corresponding photoproduction cross sections~\citep{Cepila:2017nef,Bendova:2020hbb,Mantysaari:2022sux}.

\subsection{Nuclear shadowing as a higher-twist effect}

While the Gribov-Glauber theory of nuclear shadowing and its extensions considered above are based on the space-time picture of strong interactions in the target rest frame, one can consider nuclear shadowing as an effect of small-$x$ parton recombination in the infinite momentum frame~\citep{Gribov:1983ivg,Mueller:1985wy}. In this gluon-dominated regime, nuclear shadowing is generated perturbatively through the modified scale evolution of nuclear PDFs, which in addition to the usual DGLAP form contains the higher-twist nonlinear term quadratic in the gluon density. Since the additional term is negative and enhanced in nuclei by the factor of $A/R_A^2 \sim A^{1/3}$, it slows down the $Q^2$ evolution of PDFs and, as a result, leads to significant gluon nuclear shadowing for heavy nuclei~\citep{Eskola:1993mb}.

A conceptually similar approach was considered in~\cite{Qiu:2003vd}, where one
calculates a perturbative expansion of nuclear structure functions in powers of
nucleus-enhanced higher-twist corrections proportional to $(A^{1/3}-1)/Q^2$.
Resumming these corrections, one obtains the following approximate expressions for the transverse and longitudinal nuclear structure functions,
\begin{equation}
\begin{split}
 F_T^A(x,Q^2) & \approx AF_T^N\left(x\left[1+\frac{(A^{1/3}-1) \xi^2}{Q^2}\right],Q^2\right) \,, \\
 F_L^A(x,Q^2) & \approx AF_L^N(x,Q^2)+\frac{4 \xi^2}{Q^2} F_T^A(x,Q^2) \,,
 \label{eq;shadowing_HT}
\end{split}
\end{equation}
where $\xi^2$ represents the characteristic scale, which can be expressed through the gluon distribution in the $x \to 0$ limit and $R_A$. Treating it a free parameter, one finds that $\xi^2=0.09 - 0.12$ GeV$^2$ provides a consistent description of the data on $F_2^A/F_2^d$ as a function of $x$ for light and heavy nuclei, see Fig.~\ref{fig:shadowing_dipole} (right).

\section{Global analysis approach to nuclear PDFs}
\label{sec:global-analysis}

Beyond first-principles nuclear-physics intuition and model-dependent developments, the understanding of nuclear effects in PDFs has advanced through the introduction of nuclear-PDF global analyses. The idea is to parametrise and fit the nuclear PDFs by using the available experimental data on several hard processes discussed in Sec.~\ref{sec:experiment}. This gives a data-driven approach for obtaining the nuclear PDFs for a set of nuclei. From early pioneering analyses using a small number of parameters adjusted by hand to fit DIS data at leading order in perturbative QCD, these works have evolved to truly global multi-process analyses utilising several thousand data points from different processes with (next-to)-next-to-leading order precision and using robust statistical inference tools. Today, various groups provide periodically updated and improved sets available for the use of the broader community.

In general, these analyses aim to be model agnostic, as in they do not make any a priori assumptions on what are the physics mechanisms behind the fitted nuclear effects or how they should be modelled. As such, the results of these fits cannot be straightforwardly mapped onto any single physical picture of nuclear modifications. Ultimately, one would like to constrain the fitted nuclear PDFs to a level of precision that allows for discriminating between competing theoretical models of nuclear modifications.

The global extraction of nuclear PDFs shares many details with free-proton PDF analyses, see the reviews of~\cite{Kovarik:2019xvh,Ethier:2020way} and the chapter on global analysis in this Encyclopedia. In particular, the results of these fits depend on the selection of data included in the fit, the theoretical description used for these processes, including the perturbative order and heavy-quark scheme, and further methodological choices such as the applied parametrisation of the fitted functions and the treatment of uncertainty estimation. For a more extensive discussion on the data sets included in different analyses and the methodological choices specific to the nuclear-PDF analyses, see the recent review by~\cite{Klasen:2023uqj}.

\subsection{Different parametrisation strategies}

For any PDF global analysis, a necessary starting point is to parametrise the fitted functions. A particular difference to the free-nucleon analyses in nuclear-PDF fits is the need to parametrise the dependence on both $x_N$ and $A$. In principle, also the $Z$ dependence could be parametrised directly, but typically these fits impose the bound-nucleon decomposition of Eq.~\eqref{eq:full-nucleus-from-bound-nucleons} and the isospin symmetry through Eq.~\eqref{eq:bound-nucleon-isospin} in order to simplify the isospin dependence and reduce the degrees of freedom that need to be fitted. For only a few specific nuclei is it possible to perform a single-nucleus fit with reasonable flexibility and flavour dependence in the fit functions, see~\cite{Derakhshanian:2026zkx} for the first results on the lead nucleus, where a wealth of LHC data is now available.

From early on, a prominent way to fit the nuclear PDFs has been through parametrising the nuclear modification ratios, either in terms of Eq.~\eqref{eq:nucl-mod-bound-nucleon} or~\eqref{eq:nucl-mod-full-nucleus}. These ratios then need to be multiplied with the free-nucleon PDFs to obtain the corresponding nuclear PDFs. This approach has been particularly attractive historically, since majority of the data have been available as nuclear ratios, where the dependence on the free-nucleon PDFs then largely cancels, and has been taken particularly in the analyses by \cite{Eskola:1998iy,Eskola:1998df,Eskola:2007my,Eskola:2008ca,Eskola:2009uj,Eskola:2016oht,Eskola:2021nhw}, \cite{Hirai:2001np,Hirai:2004wq,Hirai:2007sx}, and others~\citep{deFlorian:2011fp,AtashbarTehrani:2012xh,Khanpour:2016pph,Khanpour:2020zyu}. A variation on this idea appeared in~\cite{deFlorian:2003qf}, where the nuclear PDFs are defined through a convolution formula similar to Eq.~\eqref{eq:conv-form}, but trading $f^{N/A}$ with flavour dependent weight functions $W^A_i$. In the recent years, it has become popular to parametrise the bound-nucleon PDFs directly, either with an analytic functional form~\citep{Schienbein:2009kk,Kovarik:2015cma,Kusina:2020lyz,Segarra:2020gtj,Duwentaster:2021ioo,Duwentaster:2022kpv,Walt:2019slu,Helenius:2021tof} or through neural networks~\citep{AbdulKhalek:2019mzd,AbdulKhalek:2020yuc,AbdulKhalek:2022fyi}. Besides these model-agnostic fits, one can construct a formalism where the fit parameters can be more directly interpreted in terms of a certain theoretical framework, see~\cite{nCTEQ:2023cpo,Paakkinen:2025pcw} for discussion on a nucleon-nucleon short-range correlation motivated approach.

Whether one parametrises the nuclear PDFs directly or through the nuclear modification factors should not impact the outcome of the fit if sufficiently flexible parametrisation is used and the correlations with free-nucleon PDFs are properly accounted for. The latter arise from the fact that in these fits, it is necessary to introduce a free-proton baseline in order to include data which involve collisions with free protons. However, as limited amount of data is available, one generally needs to make some simplifying assumptions about the parametric dependence. Furthermore, both the $x_N$ and $A$ dependence can be parametrised equally well with various analytic functions, and for a model-agnostic fit, one cannot resort to theory to tell which parametrisation to use. For heavy nuclei, the nuclear effects tend to grow rather smoothly as the nuclear mass increases, but for light nuclei this is not necessarily the case. This can lead to a strong parametrisation dependence in regions where there are no direct data constraints. Together with the option of using either directly the cross sections or their ratios in the fit, such methodological choices can impact both the fit results for the nuclear PDFs and their correlations with the free-proton PDFs. The impact of proton-PDF uncertainties on the extraction of nuclear PDFs has been studied in the works of~\cite{Eskola:2021nhw,Eskola:2022rlm,AbdulKhalek:2019mzd,AbdulKhalek:2020yuc,AbdulKhalek:2022fyi}.

\subsection{Current status of constraints}

\begin{figure}[t]
\centering
\includegraphics[height=4.85cm]{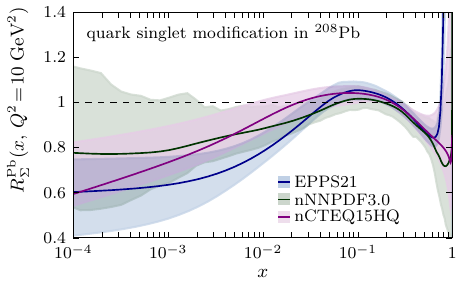}
\hspace{0.05cm}
\includegraphics[height=4.85cm]{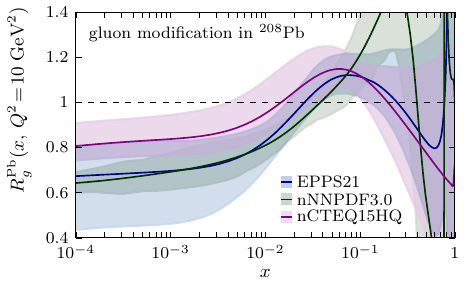}
\caption{(Left) Nuclear modification ratio for the total quark singlet in $^{208}$Pb as a function of $x_N$ at $Q^2=10$ GeV$^2$, with a comparison of the results from three recent global analyses, EPPS21~\citep{Eskola:2021nhw}, nCTEQ15HQ~\citep{Duwentaster:2022kpv}, and nNNPDF3.0~\citep{AbdulKhalek:2022fyi}. (Right) Same, but for the gluon distribution. Central best-fit results are shown as solid lines and the shaded bands correspond to uncertainty estimates at 90\% confidence level.}
\label{fig:npdfs_from_ga}
\end{figure}

Due to limitations in the available data and differences in the associated experimental uncertainties, certain (combinations of) flavours and nuclei are significantly better constrained by experimental data than others. Figure~\ref{fig:npdfs_from_ga} shows an illustrative sample of the results from the global fits, with a comparison of three recent analyses, EPPS21~\citep{Eskola:2021nhw}, nCTEQ15HQ~\citep{Duwentaster:2022kpv}, and nNNPDF3.0~\citep{AbdulKhalek:2022fyi}. These analyses are performed at the next-to-leading order in pQCD, but use some of the widest selections of different data types in the fits. On the left, we show the nuclear modification ratio for the total quark singlet $\Sigma^A(x_N, Q^2) = \sum_q \left(f_q^A(x_N, Q^2) + f_{\bar{q}}^A(x_N, Q^2)\right)$ in the lead nucleus, characterizing the effects seen in quarks on average. This combination is rather well constrained in the region of $0.01 < x_N < 0.7$ by the fixed-target DIS and DY data for a variety of nuclei, and the results from different fits are consistent with each other. The fitted nuclear PDFs exhibit clear EMC suppression, followed by some moderate antishadowing enhancement and an onset of shadowing towards smaller values of $x_N$. For $x_N < 0.01$, the available DIS data fall below the cuts of minimum $Q^2$ imposed in these fits, and the uncertainties and mutual differences in global-fit results begin to grow. Similarly, at very large momentum fractions, the constraints are limited by the applied cuts on $W$, causing the Fermi-motion region to be susceptible to parametrisation dependence and large uncertainties.

Flavour by flavour, the uncertainties in the quark sector are larger than what is obtained for the average given by the total singlet. (For more extensive comparison plots, see the corresponding publications for these analyses and the review of~\cite{Klasen:2023uqj}) This stems from the lack of sufficient constraints for pinning down the flavour separation of the nuclear PDFs. For the up versus down quark separation, a complication arises also from the fact that most of the fitted nuclei are isoscalar, i.e.\ the $u$ and $d$ quark PDFs in these nuclei are identical under the isospin symmetry, and much of the information on the possible differences in their nuclear modifications come from only a few nuclei, $^{56}$Fe and $^{208}$Pb in particular. Constraining the strange quark modifications has also proven quite difficult (see Sec.~\ref{sec:neutrino-dis} for discussion on the complications with $\nu+A$ DIS data), making it the flavour with the largest uncertainties in these fits.

For gluon modifications in the lead nucleus, shown in Fig.~\ref{fig:npdfs_from_ga} (right), these fits are consistent in finding sizeable small-$x_N$ shadowing, as well as an enhancement in the antishadowing region. These results have benefitted significantly from the availability of LHC $p+\mathrm{Pb}$ data on various processes. Despite the general agreement in the overall behaviour, the different analyses do not seem to fully agree on the amount of shadowing for gluons at small $x_N$, and also the location and height of the antishadowing peak vary considerably. These differences can be associated with the different data selections in the analyses and the fact that different data sets can pull the fit slightly into different directions. This highlights the need for more precise measurements of the key observables for gluon constraints. The need for new data is even more pressing for lighter nuclei, where the constraints for gluons are weak at best, and the current analyses are subject to a significant parametrisation bias in their results for the nuclear mass dependence of the gluon PDF.

\subsection{Relation to higher-twist phenomena}
\label{sec:global-analysis-vs-higher-twist}

In the current praxis, the majority of global analyses assume that all of the relevant nuclear effects in the fitted hard-process data can be absorbed in the nuclear PDFs, and the leading-twist contribution dominates in the perturbative region above some minimum $Q^2$ and $p_\mathrm{T}$ cuts, typically placed in the few-GeV range. An additional $W$ cut for DIS data is used to suppress higher-twist and target-mass corrections at large $x$, though including higher-twist contributions in fits of large-$x$ DIS data have been also considered explicitly~\citep{Segarra:2020gtj}. A notable exception to this general approach is the DSSZ analysis~\citep{deFlorian:2011fp}, where nuclear effects were included also in the fragmentation functions, which breaks the expectations of a leading-twist factorization picture. Since this approach resulted in a very small nuclear modifications of the gluon PDF, it seems to be disfavoured by the LHC high-$p_\mathrm{T}$ dijet data~\citep{CMS:2018jpl}, whereas global fits which use vacuum fragmentation functions in fitting the hadron production data generally agree also with the dijet measurements. Nevertheless, the interpretation of e.g.\ the $p+A$ and $d+A$ inclusive heavy-flavour and light-hadron production measurements in terms of leading-twist nuclear PDFs is not completely unambiguous. The low-$p_\mathrm{T}$ suppression in mid- and forward rapidities have been attributed alternatively (partially or entirely) to fully coherent energy loss (FCEL) or CGC effects, see Fig.~\ref{fig:inclhadron}, which involve genuine power corrections or all-twist resummations.

This emphasises the necessity of having sufficient kinematical lever-arm in the experimental data to test the validity of the leading-twist, linear DGLAP evolution picture across different values of $x_N$ and $Q^2$. Importantly, the results of the nuclear-PDF global analyses seem to comply with the picture where all the fitted data over a large kinematical phase space and a long list of processes are well described by the same universal nuclear PDFs with no additional effects involved. Any nuclear effects that are not factorisable into the initial-state nuclear PDFs, beyond-leading-twist, or process dependent should show up in these fits as tensions between different data sets. In this sense, the global analyses of nuclear PDFs can be seen also as a systematic way to search for the higher-twist effects, see~\cite{Arleo:2025oos} for discussion. The good fits to the available global data achieved in these analyses, with no significant tensions between the data types, supports the validity of collinear factorization of hard processes in lepton-nucleus and hadron-nucleus collisions and the dominance of leading-twist nuclear PDF effects in these probes over the fitted kinematic range, within the current experimental uncertainties. The possible small pulls to different directions seen in these fits, e.g.\ in the case of $p+A$ electroweak bosons versus hadronic observables and charged-lepton versus neutrino DIS, require further attention but are currently not in conflict with the general conclusion. Scrutinizing this connection and finding the limits of leading-twist dominated region remains a task for the future analyses with increasing number of and more precise data.

\section{Summary and outlook}
\label{sec:summary}

Nuclear parton distributions are indispensable quantities for understanding the structure of atomic nuclei at the fundamental, elementary particle level and have a rich phenomenology in hard nuclear scattering processes. By using coherence-length arguments, the nuclear effects in parton distributions can be classified into regions of large momentum fraction $x$, where they originate from the energy-momentum distributions and internal partonic structure of individual bound nucleons and possible non-nucleonic constituents, and small $x$, where the nuclear parton distributions must be seen as genuinely collective properties of the multi-nucleon system, not simple sums of single-nucleon contributions. The latter is the regime of nuclear shadowing, which plays an essential role in nuclear collisions at high energies, where a large fraction of the participating partons are probed at small $x$. In this chapter, we have described the basic theoretical foundation of these concepts in the context of collinear factorization and multiple coherent scattering. Through an overview of the experimental data that are used for studying the nuclear hard interactions, we showed how they paint a picture of universal, process independent, nuclear modifications of the parton distributions, and finally discussed how they are used in global analyses to perform data-driven extractions of the nuclear PDFs.

The topics of nuclear parton distributions and nuclear shadowing continue to be fields of active research. Although there is a good general theoretical understanding of the mechanisms that cause the nuclear effects seen in hard-scattering measurements, as laid out in the sections above, distinguishing between and constraining the allowed theory space of specific models appears to be difficult with the presently available data. To this end, future measurements at various experimental facilities are essential.

Some important data sets from LHC Run 2 are still not available during the writing of this chapter. In particular, $p+\mathrm{Pb}$ dijet measurements from this run are currently provided only in terms of multiplicity-binned measurements that cannot be used directly in global fits of nuclear PDFs. These data can be expected to bring important new constraints on the gluon distribution in lead over a wide region in $x$. Similarly, measurements from the Run 3 proton-oxygen data taking will bring much needed information on the $A$-dependence of the gluon PDF nuclear modifications~\citep{Paakkinen:2021jjp,Jonas:2026yoz}.

With the continuing program of UPC measurements at the LHC, one is actively exploring new observables sensitive to nuclear PDFs in the nuclear shadowing region and, in general, to small-$x$ QCD dynamics. Some recent examples in the case of Pb-Pb UPCs include incoherent $J/\psi$ photoproduction, whose  energy dependence at fixed large $|t|$ is consistent with the pattern predicted by gluon saturation models~\citep{ALICE:2025cuw}, inclusive $D^0$ photoproduction in a wide interval of the $D^0$ transverse momentum $p_T$ down to $p_T=0$~\citep{Nese:2025ohz}, and inclusive dijet photoproduction in the 0n0n neutron class, providing an observation that the modifications to nuclear PDFs vary with impact parameter~\citep{ATLAS:2026sct}.

In the future, measurements of the nuclear structure functions $F_2^A$ and $F_L^A$ in $l^{\pm}+A$ DIS for a wide range of nuclei at the EIC are expected to provide crucial tests of dynamical mechanisms of nuclear shadowing and new constraints for the nuclear PDFs over a large phase-space of $x$ and $Q^2$~\citep{Frankfurt:2002kd,Aschenauer:2017jsk}. The complications arising from tensions between neutrino-DIS data sets and from the fact that some of these data have been included in extractions of free-proton PDFs also still wait for an ultimate resolution. To this end, the possibility of measuring both charged-lepton and neutrino induced DIS at the Forward-Physics-Facility (FPF) offers an exiting opportunity to clear some of the disputes~\citep{Feng:2022inv,Francener:2025tyh}.

\begin{ack}[Acknowledgments]
\,
The research of V.~Guzey was funded by the Center of Excellence
in Quark Matter of Research Council of Finland (projects 346325 and 346326).
\end{ack}

\bibliographystyle{Harvard}
\bibliography{reference}

\end{document}